\RequirePackage{ifpdf}
\documentclass[12pt,a4paper,hyper]{JHEP3}
\usepackage{epsfig}
\usepackage{amsmath,amsfonts,amssymb}
\usepackage{multirow}
\usepackage[titletoc]{appendix}
\usepackage{slashed}
\usepackage{graphicx}
\usepackage{epsfig}
\usepackage{psfrag}
\usepackage{color}

\usepackage{ulem}
\RequirePackage{filecontents} 

\def\={\; = \;}

\font\manual=manfnt
\def\dbend{\lower3.5pt\hbox{\manual\char127}}

\def\bar{\overline}

\def\CN{{\cal N}}
\def \D{\text{D}}
\def \aD {\bar{\text{D}}}
\def \M{\text{M}}
\def \aM {\bar{\text{M}}}

\title{\center{U-duality Invariant Quantum Entropy\\ from Sums of Kloosterman Sums}}

\preprint{}
\author{ Jo\~ao Gomes\\
	
	\it Institute for Theoretical Physics, University of Amsterdam,
	Science Park 904, Postbus 94485, 1090 GL Amsterdam, The Netherlands \\ \vskip .5mm

	Email:
	\email{J.M.VieiraGomes at uva.nl}
}

\abstract{U-duality plays a special role in the study of the microscopic degrees of freedom of supersymmetric black holes. To be consistent with duality, the black hole quantum degeneracy must  obey special arithmetic properties, which are non-perturbative in nature.  In this work, we study these properties from a holographic point of view, establishing a connection between arithmetic properties of Kloosterman sums and quantum gravity in $\text{AdS}_2$ space. To this end, we consider the entropy of black holes that carry non-primitive charges, in both $\CN=8$ and $\CN=4$ four dimensional compactifications; our analysis includes all the perturbative and non-perturbative bulk quantum corrections. The key result relies on special arithmetic properties of generalized Kloosterman sums that we develop. These are a generalization of the known Selberg identity of classical Kloosterman sums. In both the $\CN=8$ and $\CN=4$ examples, we recover, from the bulk quantum gravity, the non-primitive answer which is a sum over the primitive degeneracies, depending non-trivially on the discrete duality invariants. In particular, for the $\CN=4$ case we show that the quantum gravity answer reproduces the dependence on the torsion invariant $I=\text{gcd}(Q\wedge P)$, in agreement with the microscopic formulas. For the $\CN=8$ case, we solve a puzzle related to U-duality invariance of the supergravity answer and the corresponding one-eighth BPS degeneracy.   }

\keywords{holography, supergravity, Localization}

\begin{document}
\section{Introduction}

U-duality \cite{Hull:1994ys,Sen:1994fa,Sen:1998kr} is one of the most important concepts that stems from string theory. It has played a key role in our understanding of the spectrum of BPS states in string theory and quantum field theories. Since it is a non-perturbative map between different string theories, we can use the duality to count BPS states in a frame where known field theory methods can be applied.  This has been extensively used  to explain the statistical origin of the entropy of supersymmetric black holes \cite{Strominger:1996sh}.

In this work we look at U-duality from a holographic point of view. We are interested in understanding what are the implications of duality for quantum black holes, in particular, to the structure of quantum corrections to the entropy, beyond the leading area formula. Our focus is to understand how the bulk quantum gravity, namely the path integral over the string fields on the near-horizon geometry, can explain the discrete non-perturbative structure that U-duality imposes  on black hole entropy. 
 
The index that counts  BPS states is an invariant under U-duality, and so it is expected to be a function of duality invariants only; it is a highly non-trivial problem to determine how the index  depends on these invariants. However, for a large class of supersymmetric configurations in $\CN=8$ and $\CN=4$ compactifications to four dimensions, there is partial understanding on the dependence of the index on discrete arithmetic U-duality invariants \cite{Maldacena:1999bp,Sen:2008sp,Banerjee:2008pu,Dabholkar:2008zy,Banerjee:2008pv}. Such dependence is usually characterized by a sum of the form
\begin{equation}\label{non-primitive deg}
d_I(\Gamma)=\sum_{s|I}g(s)\,d_{I=1}(\Gamma/s).
\end{equation}
Here $I$ denotes the discrete invariants, $g(s)$ is an arithmetic function, with the condition that $g(1)=1$, and $\Gamma/s$, which is integer, denotes a particular rescaling of the charge vector $\Gamma$ by $s$. The index $d_I(\Gamma)$ is the non-primitive degeneracy and $d_{I=1}$ is the primitive answer, with primitivity being associated with whether the discrete invariant is one or not.  Later we will describe in detail what is $g(s)$ for the $\CN=8$ and $\CN=4$ examples.

For these supersymmetric compactifications there is an equality between index and degeneracy \cite{Sen:2009vz,Dabholkar:2010rm}, which we can use to extract precise data for black hole entropy. From the non-primitive formula (\ref{non-primitive deg}), the sum over the divisors $s|I$  gives rise to non-perturbative corrections to the leading black hole entropy. That is, for large charges each of the $d_{I=1}$ has exponential growth, in agreement with the Bekenstein-Hawing entropy area law \cite{Bekenstein:1973ur}, and hence we can approximate
\begin{equation}\label{Z_s orbifolds}
d_{I=1}(\Gamma/s)\simeq \exp{\left(\frac{A}{4 s}\right)},\;A\gg 1,
\end{equation}
 where $A/4$ is the area formula, which depends on the charges $\Gamma$ in a particular way. The leading term in (\ref{non-primitive deg}) is thus the term with $s=1$, while the term with $s=I$ is the most subleading. We thus see that the terms with $s> 1$ are exponentially suppressed relative to the term with $s=1$. 

In the context of the quantum entropy and the $\text{AdS}_2/\text{CFT}_1$ correspondence \cite{Sen:2008vm}, it was proposed in \cite{Sen:2009vz,Sen:2009gy} that such non-perturbative corrections could be understood from orbifold geometries of the form $AdS_2\times S^2\times S^1/\mathbb{Z}_s$ \footnote{We are omitting details about the compactification manifold. We will make this clear later on.} in the $AdS_2$ path integral. The sum over the divisors $s|I$ appeared as a result of a smoothness condition of the three-form fluxes in IIB string theory, which parametrize the electric and magnetic charges in this frame. Due to the orbifold, the large charge contribution of these new saddles to the quantum entropy is the exponential of the classical area divided by $s$, precisely as predicted by the formula (\ref{Z_s orbifolds}). Nevertheless, in \cite{Sen:2009vz,Sen:2009gy} it remained an assumption whether the full quantum entropy on each of the $\mathbb{Z}_s$ orbifold saddles gave the primitive $d_{I=1}$ formula with rescaled charges, as in the non-primitive answer (\ref{non-primitive deg}). To put it in a different way, \cite{Sen:2009vz,Sen:2009gy} could not explain why the path integral on each of the orbifold saddles did not depend on the residual arithmetic properties of the charges but only on the T-duality invariants, which parametrize uniquely the U-duality orbits in the primitive case. The puzzle we want to answer is how the $AdS_2$ path integral reduces, for general charge configurations, to the structure (\ref{non-primitive deg}), as a sum over the primitive formulas.

On the other hand, recent work on non-perturbative corrections to black hole entropy using localization techniques \cite{Dabholkar:2014ema,Gomes:2017bpi}, motivated by the analysis of \cite{Murthy:2009dq}, has provided evidence that the quantum entropy path integral in M-theory receives the contribution of $AdS_2\times S^2\times S^1/\mathbb{Z}_c$ orbifolds with arbitrary values of $c\geq 1$. These orbifolds also give rise to corrections to the entropy of the form 
\begin{equation}\label{Z_c orbifolds}
\sim \exp{\left(\frac{A}{4 c}\right)},\;A\gg 1,
\end{equation}
where the $1/c$ factor is due to the orbifold. 

From the M-theory point of view, one also expects corrections of the form (\ref{Z_s orbifolds}), because the index is invariant under duality. We may wonder if such corrections are related to the contribution of $\mathbb{Z}_s$ orbifolds in the path integral, as proposed in the IIB frame \cite{Sen:2009vz,Sen:2009gy}. This possibility, however, seems to be in tension with the fact that we can sum over arbitrary $\mathbb{Z}_c$ orbifolds in the path integral, that is, with no restriction on $c$.  Furthermore, if we are summing over arbitrary $\mathbb{Z}_c$ orbifolds then it is puzzling to understand how the path integral can distinguish between corrections of the form $\exp(A/4c)$ over $\exp(A/4s)$, with $s$ dividing the arithmetic invariant $I$.

The goal of this paper is to provide a solution to this puzzle, making the two types of non-perturbative corrections, (\ref{Z_s orbifolds}) and (\ref{Z_c orbifolds}), compatible with each other. We will show from the  M-theory point of view that the path integral naturally reproduces  the degeneracy as a sum over the primitive degeneracies $d_{I=1}$ as in (\ref{non-primitive deg}), including the dependence on the duality invariants for both the $\CN=8$ and $\CN=4$ black holes. Nevertheless, it will fail to reproduce exactly the arithmetic function $g(s)$, and we comment upon this, providing with a possible solution.  Our analysis is exact and thus includes all the perturbative and non-perturbative corrections to the area formula. To achieve this we will use recent developments using supersymmetric localization techniques \cite{Dabholkar:2010uh,Dabholkar:2011ec,Gupta:2012cy}.

At the heart of our solution is the recent proposal for a non-perturbative description of black hole entropy \cite{Gomes:2017eac}. Our results will serve as a test to that construction. This proposal attempts a first principles derivation of non-perturbative effects in the $AdS_2$ path integral, associated to the wrapping of $\M2$  and $\aM2$-branes on cycles of the Calabi-Yau manifold- this is the effect of integrating out branes. In this proposal, the M-theory path integral on the near-horizon geometry of the black hole receives the contribution of additional saddle geometries, which are conjectured to describe the physics near the core \footnote{This obtained after a sensible decoupling limit from the asymptotic physics.} of a $r$ Taub-Nut (TN) and $r$ anti-Taub-Nut ($\bar{\text{TN}}$) geometry, which is the uplift of a pair of  $r$ $\D6$ and $r$ $\aD6$ configuration in type IIA wrapping the Calabi-Yau directions, with singular G-fluxes\footnote{The G-flux corresponds to the M-theory four-form field strength.} turned on the Calabi-Yau. In \cite{Gomes:2017eac}, the focus was mainly on solutions with $r=1$, but here we argue that in order to describe the arithmetic properties of black hole entropy we need to consider $r>1$. In this case, the geometry becomes an $AdS_2\times S^1\times S^2/\mathbb{Z}_r$ orbifold which is not freely acting and so has singularities at the fixed points of rotations, that is, at the origin of $AdS_2$ and at the north and south poles of $S^2$. The reason we include this singular geometry in the path integral is because the $\D6$ and $\aD6$ branes, from the IIA perspective, are sitting respectively at those fixed points, which we can take as the physical regulators. At the moment, we do not understand how these singularities are resolved, but we will show that this construction leads to the desired result.

From our perspective, the path integral receives the contribution of $AdS_2\times S^1\times S^2 /\mathbb{Z}_c$ orbifolds with arbitrary $c$. The orbifold acts on both the $U(1)$ isometries of $AdS_2$ and $S^2$, but also identifies points on the circle $S^1$ as $y\sim y+2\pi R d/c$ with $0\leq d<c$, with $R$ the radius. The key feature of our solutions, in contrast with \cite{Murthy:2009dq,Dabholkar:2014ema,Gomes:2017bpi}, is that we relax the condition that $d$ and $c$ are relatively co-prime. In this case, the orbifold can be singular if $d$ and $c$ have common factors. To be consistent with the $\D6-\aD6$ picture, or equivalently the Taub-Nut/anti-Taub-Nut geometry, we allow for geometries which have $\text{gcd}(c,d)\leq r$, and so the orbifold singularities are at most of $\mathbb{Z}_r$ type. In the path integral we also include smooth geometries which we can see as $s$-covers of the $\mathbb{Z}_s$ orbifolds, with $s|r$; this happens when $(c,d)$ are co-prime.  In addition, the $\mathbb{Z}_c$ orbifolds are accompanied by a change in the homology of the contractible and non-contractible cycles  inside  $AdS_2\times S^1$, which is topologically a solid torus. This results in the sum over the $M_{(c,d)}$ geometries studied in \cite{Murthy:2009dq,Dabholkar:2014ema} but now with the condition $\text{gcd}(c,d)\leq r$. The homology change of the contractible and non-contractible cycles is characterized by a two dimensional matrix of determinant $r$, which determines how we fill in the solid torus. Our task is to reproduce the results of \cite{Dabholkar:2014ema,Gomes:2017bpi} for the full quantum entropy including the Kloosterman sums in these new geometries.

The answer for the path integral on each $M_{(c,d)}$ geometry turns out be a Bessel function multiplied by a modified version of the Kloosterman sum that we denote momentarily by $\tilde{K}l(m,n,p,c,r)$, where $(m,n,p)$ are the black hole charges; the $(m,n,p)$ charges turn out to be related to the T-duality invariants. This modified Kloosterman sum differs from the usual definition by the fact that  one sums over integers $a,d$ that obey $ad=r\text{ mod}(c)$, which is the condition that the matrix that determines the homology cycles has determinant $r$.   When the black hole charges $(m,n,p)$ have common factors, we can show that the modified Kloosterman sums have non-trivial arithmetic properties, characterized by a sum over Kloosterman sums. These sums turn out to depend on the discrete U-duality invariants as predicted from the microscopic formulas. After some algebra the full degeneracy naturally acquires the non-primitive form (\ref{non-primitive deg}).

The arithmetic properties we will be exploring  are a generalization of the Selberg identity  of classical Kloosterman sums \cite{Selberg}. This identity is of the form
\begin{equation}\label{Selberg id}
Kl(m,n,c)=\sum_{s|(m,n,c)}s\, Kl(mn/s^2,1,c/s),
\end{equation}
where $Kl(m,n,c)$ are the classical Kloosterman sums
\begin{equation}
Kl(m,n,c)=\sum_{\substack{a,d\in \mathbb{Z}/c\mathbb{Z}\\ ad=1\text{ mod}(c)}} \exp{\left[2\pi i m \frac{d}{c}+2\pi i n \frac{a}{c}\right]}.
\end{equation}
 A derivation of the formula (\ref{Selberg id}) was given in \cite{Kuznecov,Matthes1990} \footnote{The formula (\ref{Selberg id}) was stated by A. Selberg without proof.}. One of the main results of this work is the derivation of similar arithmetic properties for generalized Kloosterman sums, which include multiplier matrices for Jacobi forms of arbitrary index \cite{Dijkgraaf:2000fq}. These properties will show to be crucial to obtain the structure of the non-primitive degeneracy formula (\ref{non-primitive deg}).

The analysis that we present in this work will allow us to solve an important puzzle related to the one-eighth BPS black hole degeneracy. The puzzle is related to the fact that, in this case, the microscopic formula depends only on the data of the CFT for one $\D1$ and one $\D5$-brane (from the type IIB perspective), so the data of a system with central charge $c=6$. However, from supergravity we may have different central charges, depending on which magnetic charges we pick. In particular, these magnetic charges determine the different Chern-Simons levels of the effective three dimensional theory dual to the CFT \cite{Kraus:2005vz,Dabholkar:2010rm}, and hence play a crucial role in the computation of the Kloosterman sums from the bulk theory \cite{Dabholkar:2014ema,Gomes:2017bpi}. For example, to match the computation of the Kloosterman sums  against the microscopic formula in \cite{Dabholkar:2014ema}, the authors impose a particular choice of charges, namely that the central charges are those of a single $\D1$ and $\D5$-brane configuration. However, to make contact with weakly coupled physics in the bulk theory, we require large central charge, which is clearly in tension with what we have just described. So there seems to be an apparent contradiction between the gravity picture  and the microscopic formulas. Our results solve this problem. On one hand, we show that the gravity theory admits a weakly coupled regime provided that $r\gg 1$, and on the other hand, we show that it reduces to the data of the CFT with $c=6$ using the properties of Kloosterman sums.

The plan of the paper is as follows. In the section \S \ref{sec micro deg} we review the microscopic degeneracy formulas for dyons with non-primitive charges for both $\CN=8$ and $\CN=4$ four dimensional compactifications. Then in section \S \ref{sec Selberg} we discuss and derive arithmetic properties of classical and generalized Kloosterman sums. We will use two different but equivalent methods: one uses the properties of Hecke operators and the other is purely algebraic. In section \S \ref{sec hol comp} we consider the holographic bulk computation of the entropy on the new geometries, which include the singular $\mathbb{Z}_r$ cases.  We will find that the path integral reproduces exactly the Bessel function multiplied by a modified Kloosterman sum. We then use the arithmetic properties of these Kloosterman sums developed in section \S \ref{sec Selberg} to show that the full quantum degeneracy becomes a sum over the primitive formulas, carrying non-trivial dependence on the arithmetic U-duality invariants. Finally in section \S \ref{sec Hecke} we address the problem of writing the non-primitive degeneracy in the form of an Hecke operator acting on the primitive formulas. We will argue this can be used to generate additional dependence on other duality invariants.

\section{Microscopic Degeneracy of Non-Primitive Dyons}\label{sec micro deg}

In this section we review the microscopic degeneracy formulas for non-primitive dyons in $\CN=8$ \cite{Maldacena:1999bp,Sen:2008sp} and $\CN=4$  compactifications \cite{Banerjee:2008pu,Dabholkar:2008zy}. The formulas that we describe, represent, nonetheless, only a subset of the full U-duality symmetry. The dependence of the microscopic formulas on the full set of U-duality invariants is still unknown. 

\subsection{One-eighth BPS Dyons in $\CN=8$}\label{sec non-primitive N=8}

We recall the U-duality invariant formula proposed in \cite{Sen:2008sp} which follows from the original work of \cite{Maldacena:1999bp}. To facilitate the analysis, the charge vectors are described with respect to a $SL(2,\mathbb{Z})\times SO(6,6,\mathbb{Z})$ subgroup of the U-duality group $E_{7,7}(\mathbb{Z})$ of IIB on $T^6$. The fundamental representation of $E_{7,7}(\mathbb{Z})$ decomposes as $\mathbf{56}=(\mathbf{2},\mathbf{12})+(\mathbf{1},\mathbf{32})$. Physically, the $(\mathbf{2},\mathbf{12})$ charges correspond respectively to the electric and magnetic $Q^i$ and $P^i$ charges in the NSNS sector, with $i=1\ldots 12$; the $SL(2,\mathbb{Z})$ factor acts as an electric-magnetic symmetry. The remaining $(\mathbf{1},\mathbf{32})$ are Ramond-Ramond charges. The U-duality formula of \cite{Sen:2008sp} assumes charge vectors that are purely NSNS or that can be brought to such a configuration by an $E_{7,7}(\mathbb{Z})$ transformation, and so we always assume that the $(\mathbf{1},\mathbf{32})$ are absent.

Define the charge combinations
\begin{equation}
I_1=\text{gcd}(Q_iP_j-Q_jP_i),\;\;I_2=\text{gcd}\left(Q^2/2,P^2/2,Q.P\right).
\end{equation}
Here $Q^2, \,P^2, \, Q.P$ are the T-duality invariants which are constructed out of bilinears of the form $Q^i L_{ij}Q^j$, $P^iL_{ij}P^j$ and $Q^iL_{ij}P^j$ respectively, with $L_{ij}$ a metric invariant under $SO(6,6,\mathbb{Z})$. The $I_1$ and $I_2$ combinations are individually arithmetic invariants under $SL(2,\mathbb{Z})\times SO(6,6,\mathbb{Z})$, but under a general $E_{7,7}(\mathbb{Z})$ transformation $I_1$ and $I_2$ are not left invariant. What is invariant is the combination $\text{gcd}(I_1,I_2)$. In fact, such combination can be written as $\text{gcd}(q\otimes q)_{\textbf{133}}$, where $q$ is the $ \textbf{56}$ representation of $E_{7,7}(\mathbb{Z})$,  and the $\textbf{133}$ susbcript denotes the $\textbf{133}$ dimensional representation. Following \cite{Sen:2008sp}, in this work we consider only the case with
\begin{equation}\label{primitivity cond 1}
\psi(q)=\text{gcd}(q\otimes q)_{\textbf{133}}=1 \Leftrightarrow \text{gcd}(I_1,I_2)=1.
\end{equation}
Later we will comment on the generalization of this invariant for non-primitive values. The degeneracy formula proposed in \cite{Sen:2008sp} for this class of charge vectors is
\begin{equation}\label{duality deg}
d(Q,P)=(-1)^{Q.P}\sum_{s|I_1I_2}\, s\,c\left(Q^2P^2/4s^2, Q.P/s\right),\;\text{gcd}(I_1,I_2)=1.
\end{equation}
The coefficients $c(n,l)$ are the Fourier coefficients of the index one Jacobi form 
\begin{equation}\label{Jacobi N=8}
\frac{\vartheta^2(\tau,z)}{\eta^6(\tau)}=\sum c(n,l)q^ny^l,\;\;q=e^{2\pi i\tau},\,y=e^{2\pi i z},
\end{equation}
with $\vartheta(\tau,z)$ the odd Jacobi theta function and $\eta(\tau)$ the Dedekind function. When $I_1=1$ the expression (\ref{duality deg}) reproduces the formula derived originally in \cite{Maldacena:1999bp}. Since the Jacobi form (\ref{Jacobi N=8}) has index one, the coefficients $c\left(Q^2P^2/4s^2, Q.P/s\right)$ depend only on the quartic charge combination $\Delta=Q^2P^2-(Q.P)^2$, which is invariant under the continuous $SL(2)\times SO(6,6)$ group.

\subsection{One-quarter BPS Dyons in $\CN=4$}

Heterotic string theory compactified on $T^6$ has U-duality group $SL(2,\mathbb{Z})\times O(6,22,\mathbb{Z})$, where the first factor acts as an electric-magnetic duality and the second is the T-duality group. A generic dyon in this theory is labelled by a pair of $28$ dimensional vectors $(Q,P)$ living in the Narain lattice $\Lambda^{6,22}$. Each $(Q^i,P^i)$ transforms as a doublet in the fundamental of $SL(2,\mathbb{Z})$ which acts as an electric-magnetic duality transformation; the charges $Q$ and $P$ are therefore the electric and magnetic charges respectively. Besides, each of the charge vectors transforms in the vector representation of the T-duality group $O(6,22,\mathbb{Z})$. 

 A generic $(Q,P)$ charge configuration can be brought to the form \cite{Banerjee:2008ri}
 \begin{equation}
 (Q,P)=(IQ_0,P_0),
 \end{equation}
  with $I\in \mathbb{Z}$, and it has the property that $\text{gcd}(Q_0\wedge P_0)=1$.  This also means that the pair $(Q_0,P_0)$ lies along primitive vectors in the Narain lattice. The integer $I$ therefore equals 
 \begin{equation}
 \text{gcd}(Q\wedge P)=I,
 \end{equation}
 which is known as torsion and it is invariant under the discrete U-duality group \cite{Dabholkar:2007vk}. 

A proposal for the microscopic degeneracy of dyons with  torsion $>1$ was put forward in \cite{Banerjee:2008pu,Dabholkar:2008zy}, though a first principles derivation is still an open problem. The proposed formula has the form
\begin{equation}\label{non-primitive N=4}
d(Q,P)=(-1)^{Q.P+1}\sum_{s|I}s\;d_{I=1}(Q^2/s^2,P^2,Q.P/s),
\end{equation}
where $d_{I=1}$ is the degeneracy for a dyon with unit torsion, also known as primitive dyon. This degeneracy can be extracted from the Fourier coefficients of the reciprocal of the weight ten Siegel modular form \cite{Dijkgraaf:1996it,David:2006yn}, that is,
\begin{equation}
\sum_{m,n,p}d_{I=1}(n,m,r)p^mq^ny^r=\frac{1}{\Phi_{10}(\tau,\sigma,z)},
\end{equation}
with $\Phi_{10}(\tau,\sigma,z)$ the Igusa cusp and $q=\exp{(2\pi i\tau)}$, $p=\exp{(2\pi i \sigma)}$ and $y=\exp{(2\pi i z)}$. The primitive counting can be generalized for one-quarter BPS dyons in $\CN=4$ CHL compactifications, which are $K3$ and $T^4$ orbifold compactifications. The canonical partition function in this case is the reciprocal of a Siegel modular form of a congruence subgroup. Though our analysis of the bulk entropy can be extended to these CHL examples, our focus will be on the dyons captured by the Igusa form.

\section{Selberg Identities and Sums of Kloosterman Sums}\label{sec Selberg}

The  Selberg identity of classical Kloosterman sums is the identity (\ref{Selberg id}). In this section we re-derive that equality using two different methods. The first uses the properties of the Hecke operators acting on modular forms, and the Rademacher expansion of the Fourier coefficients. This method is very similar to the one used in \cite{Kuznecov}. The second method is purely algebraic and consists in constructing a modified version of the classical Kloosterman sums. We can then show that the modified versions have  arithmetic properties that encode the Selberg formula.  In the second part of this section we use the same methods to produce new arithmetic identities for the generalized Kloosterman sums, which appear in the Rademacher expansion of the Fourier coefficients of Jacobi forms.

\subsection{Classical Kloosterman Sums}

The first method follows an idea originally used by Kuznetsov \cite{Kuznecov} to obtain the Selberg identity of classical Kloosterman sums. Our method consists in using the action of the Hecke operator on non-positive weight modular forms and their Fourier coefficients. By comparing the Rademacher expansions of the Fourier coefficients of the original modular form and their image under the Hecke operator action, we  obtain arithmetic identities for the classical Kloosterman sums. 

To exemplify our procedure we re-derive the Selberg identity (\ref{Selberg id}).  To do this we need the action of the Hecke operator $T_m$ on modular forms. Suppose $\phi_{\omega}(\tau)$ is a modular form of weight $\omega$. The Hecke operator acting on $\phi_{\omega}(\tau)$ generates a new modular form of the same weight that we denote by $\tilde{\phi}_{\omega}(\tau)=T_m\circ\phi_{\omega}(\tau)$. Following \cite{Zagier1992}, the Fourier coefficients $\tilde{c}_m(n)$ of $\tilde{\phi}_{\omega}(\tau)$ are determined in terms of the Fourier coefficients $c(n)$ of $\phi(\tau)$ as
\begin{equation}\label{Hecke on mod form}
\tilde{c}_m(n)=\sum_{s|(n,m)}s^{\omega-1}c(nm/s^2).
\end{equation}
If $\phi_{\omega}(\tau)$ is a modular form of non-positive weight that contains polar terms, that is, if diverges at $i\infty$, then the Fourier coefficients of both $\tilde{\phi}_{\omega}(\tau)$ and $\phi_{\omega}(\tau)$ can be written in a Rademacher expansion \cite{Rademacher-1938,Zuckerman-Rademacher}. For the Fourier coefficients of the modular form $\phi_{\omega}(\tau)$, the Rademacher expansion is
\begin{equation}
c(n)=\sum_{n_p<0}c(n_p)\sum_{c=1}^{\infty}\frac{1}{c}Kl(n,n_p,c)\int_{\epsilon-i\infty}^{\epsilon+i\infty}\frac{dt}{t^{2-\omega}}\exp{\left[2\pi\frac{ n}{ct}-2\pi\frac{n_p t}{c}\right]},
\end{equation}
where $Kl(n,n_p,c)$ is the classical Kloosterman sum and $c(n_p)$ is the Fourier coefficient of the polar term parametrized by $n_p$. Similarly for $\tilde{\phi}_{\omega}(\tau)$ we have
\begin{equation}\label{Rademacher 2}
\tilde{c}_m(n)=\sum_{\tilde{n}_p<0}\tilde{c}_m(\tilde{n}_p)\sum_{c=1}^{\infty}\frac{1}{c}Kl(n,\tilde{n}_p,c)\int_{\epsilon-i\infty}^{\epsilon+i\infty}\frac{dt}{t^{2-\omega}}\exp{\left[2\pi\frac{n}{ct}-2\pi\frac{\tilde{n}_p t}{c}\right]},
\end{equation}
where $\tilde{n}_p$ are the polar terms of $\tilde{\phi}_{\omega}(\tau)$. 

We now show that the Rademacher expressions of both $c(n)$ and $\tilde{c}_m(n)$, together with the action of the Hecke operator (\ref{Hecke on mod form}), give rise to special arithmetic properties of the Kloosterman sums. First we use the equality (\ref{Hecke on mod form}) to relate the  polar terms of $\tilde{\phi}_{\omega}(\tau)$ to the polar terms of $\phi_{\omega}(\tau)$. The polar terms $\tilde{n}_p$ are determined as 
\begin{eqnarray}
&&\tilde{n}_pm/s^2=n_p,\quad s|(\tilde{n}_p,m)\\
\Leftrightarrow &&\tilde{n}_p=n_pm/d^2,\quad d|(n_p,m),
\end{eqnarray}
and the polar coefficients are  related as
\begin{equation}
\tilde{c}_m(\tilde{n}_p)=\sum_{\substack{d,n_p\\ d|(m,n_p)\\ \tilde{n}_p=n_pm/d^2}} \left(\frac{m}{d}\right)^{\omega-1}c(n_p).
\end{equation}
Plugging this equation back in (\ref{Rademacher 2}) we obtain
\begin{equation}
\tilde{c}_m(n)=\sum_{n_p<0}c(n_p)\sum_{d|(m,n_p)}\left(\frac{m}{d}\right)^{\omega-1}\sum_{c=1}^{\infty}\frac{1}{c}Kl(n,n_pm/d^2,c)\int_{\epsilon-i\infty}^{\epsilon+i\infty}\frac{dt}{t^{2-\omega}}\exp{\left[2\pi\frac{n}{ct}-2\pi\frac{n_p m t}{d^2c}\right]}.
\end{equation}
Then we rescale $t$ as $t\rightarrow td/m$ to obtain
\begin{equation}\label{Rademacher 3}
\tilde{c}_m(n)=\sum_{n_p<0}c(n_p)\sum_{d|(m,n_p)}\sum_{c=1}^{\infty}\frac{1}{c}Kl(n,n_pm/d^2,c)\int_{\epsilon-i\infty}^{\epsilon+i\infty}\frac{dt}{t^{2-\omega}}\exp{\left[2\pi\frac{nm}{cdt}-2\pi\frac{n_pt}{cd}\right]}.
\end{equation}
We want to exchange the sum over $d$ with the sum over $c$. To simplify the problem consider the following sum
\begin{equation}
 \sum_{c=1}^{\infty}\sum_{s|(c,I)}s^a\,f(n^i/s,c/s),
\end{equation}
with $n^i$ an array of integers, and $f(n^i,c)$ an arbitrary function. We have denoted $I=\text{gcd}(n^i)$ and $a$ is some arbitrary integer coefficient. 
We can easily  see that the following equality holds
\begin{equation}\label{sum c sum s}
\sum_{c=1}^{\infty}\sum_{s|(c,I)}s^a\,f(n^i/s,c/s)=\sum_{s|I}s^a\sum_{c=1}^{\infty}f(n^i/s,c).
\end{equation}
To show this, it is easier to start from the RHS of equation (\ref{sum c sum s}) and write
\begin{equation}
\sum_{s|I}s^a\sum_{c=1}^{\infty}f(n^i/s,c)=\sum_{s|I}\sum_{\substack{c'=1 \\ s|c'}}^{\infty} s^a f(n^i/s,c'/s) =\sum_{c'=1}^{\infty}\sum_{s|(c',I)}s^a\,f(n^i/s,c'/s).
\end{equation}
 Though a simple expression, we will use this equality extensively throughout this work, so we decided to name it \textit{divisor sum rule}.
 
Hence, we can use the divisor sum rule (\ref{sum c sum s}) to exchange the sums over $d$  and $c$ in (\ref{Rademacher 3}) to obtain
\begin{equation}\label{Rademcher 4}
\tilde{c}_m(n)=\sum_{n_p<0}c(n_p)\sum_{c=1}^{\infty}\frac{1}{c}\sum_{d|(m,n_p,c)}d\,Kl(n,n_pm/d^2,c/d)\int_{\epsilon-i\infty}^{\epsilon+i\infty}\frac{dt}{t^{2-\omega}}\exp{\left[2\pi\frac{nm}{ct}-2\pi\frac{n_pt}{c}\right]}.
\end{equation}
Since we can also express $\tilde{c}_m(n)$ in the form $\tilde{c}_m(n)=\sum_{s|(n,m)}s^{\omega-1}c(nm/s^2)$, if we plug back in this formula the Rademacher expansion of $c(n)$, then by comparing with the expression (\ref{Rademcher 4}) we obtain arithmetic properties for the Kloosterman sums. In particular, we will show that the following property holds
\smallskip
\begin{equation}\label{main Kloos Selberg formula}
\sum_{d|(m,n_p,c)}d\,Kl(n,n_pm/d^2,c/d)=\sum_{s|(n,m,c)}s\,Kl(nm/s^2,n_p,c/s).
\end{equation}
\smallskip
To show this holds, we plug the above expression back in formula (\ref{Rademcher 4}), and rescale $t$ as $t\rightarrow ts$, to obtain
\begin{equation}
\tilde{c}_m(n)=\sum_{n_p<0}c(n_p)\sum_{c=1}^{\infty}\sum_{s|(n,m,c)}\frac{s^{\omega-1}}{(c/s)}\,Kl(nm/s^2,n_p,c/s)\int_{\epsilon-i\infty}^{\epsilon+i\infty}\frac{dt}{t^{2-\omega}}\exp{\left[2\pi\frac{nm/s^2}{(c/s)t}-2\pi\frac{n_pt}{c/s}\right]}.
\end{equation}
Using the divisor sum rule (\ref{sum c sum s}), we get
\begin{eqnarray}
\tilde{c}_m(n)&=&\sum_{s|(n,m)}s^{\omega-1}\,\sum_{n_p<0}c(n_p)\sum_{c=1}^{\infty}\frac{1}{c}\,Kl(nm/s^2,n_p,c)\int_{\epsilon-i\infty}^{\epsilon+i\infty}\frac{dt}{t^{2-\omega}}\exp{\left[2\pi\frac{nm/s^2}{ct}-2\pi\frac{n_pt}{c}\right]}\nonumber\\
&=&\sum_{s|(n,m)}s^{\omega-1}c(mn/s^2),
\end{eqnarray}
where to obtain the second line we used the Rademacher expansion for $c(n)$. We obtain (\ref{Hecke on mod form}) as we wanted to show.

From equation (\ref{main Kloos Selberg formula}), it is easy to show by setting $n_p=1$, that we must have
\smallskip
\begin{equation}
Kl(n,m,c)=\sum_{s|(n,m,c)}s\,Kl(nm/s^2,1,c/s).
\end{equation}
\smallskip
This is the Selberg identity (\ref{Selberg id}). As a consistency check, we can apply this identity to both sides of (\ref{main Kloos Selberg formula}). We get
\begin{equation}
\sum_{\substack{s=ab\\a|(m,n_p,c)\\b|(n,n_pm/a^2,c/a)}} s N(s) Kl\left(\frac{n mn_p}{s^2},1,\frac{c}{s}\right)=\sum_{\substack{s'=a'b'\\a'|(n,m,c)\\ b'|(nm/a'^2,n_p,c/a')}}s'N(s')\,Kl\left(\frac{nmn_p}{s'^2},1,\frac{c}{s'}\right),
\end{equation}
where $N(s)$ is the number of ways of writing $s=ab$, and similarly for $N(s')$. We show that the sums in both sides of the above equation are equal. Following \cite{Eichler:1985ja}, we write $a=s\delta/(s,n_p)$ for some integer $\delta$. Then we have $b=(s,n_p)/\delta$. Plugging this back in the conditions we find
\begin{equation}\label{number divisors}
N(s)=\text{number of divisors }\delta\,\left(n,m,n_p,s,\frac{nn_p}{s},\frac{nm}{s},\frac{mn_p}{s},\frac{nmn_p}{s^2}\right),
\end{equation}
with the condition that $N(s)=0$ unless $s|(nm,mn_p,nn_p)$ and $s^2|nmn_p$. Repeating the same exercise for $N(s')$,  we find again (\ref{number divisors}), which shows the equality. 

In the following, we use a different method to derive the identity (\ref{main Kloos Selberg formula}). This method will be particularly useful  later on to understand the physical origin of the duality invariant quantum black hole entropy. To do so, we consider a modified version of the classical Kloosterman sum that we define as
\begin{equation}\label{modified Kloos}
\tilde{Kl}(n,m,c,r)=\sum_{\substack{0 \leq a,d<c\\ ad=r\text{ mod}(c)}}e^{2\pi in\frac{d}{c}+2\pi im \frac{a}{c}}.
\end{equation} 
For $r=1$ we recover the classical Kloosterman sum. 

Following \cite{Zagier1992}, any $2\times 2$ matrix with determinant $r$ can be decomposed as 
\begin{equation}
\left(\begin{array}{cc}
a & b\\
c& d
\end{array}\right)=\left(\begin{array}{cc}
r/s & b''\\
0 & s
\end{array}\right)\left(\begin{array}{cc}
a' & b'\\
c' & d'
\end{array}\right), \;\text{with }\left(\begin{array}{cc}
a' & b'\\
c' & d'
\end{array}\right)\in PSL(2,\mathbb{Z})\;\text{and }s|(c,r),
\end{equation}
with $c>0$. This "left" decomposition preserves the ratios $d/c=d'/c'$. We could  have similarly considered the "right" decomposition which preserves instead the ratios $a/c=a'/c'$. We have denoted left and right representations whenever the matrix $\left(\begin{array}{cc}
* & *\\
0 & *
\end{array}\right)$ appears on the left or right side of the $PSL(2,\mathbb{Z})$ matrix respectively. On the other hand since we are quotienting on the left by $\left(\begin{array}{cc}
1 & l\\
0& 1
\end{array}\right)$ with $l\in \mathbb{Z}$, we can choose $0\leq b''<s$. Therefore, the sum over $a,d$ in (\ref{modified Kloos}) can be traded by a sum over $s|(c,r)$, $0\leq b''<s$ and $0\leq a',d'<c$. Plugging this decomposition back in (\ref{modified Kloos}) we obtain
\begin{eqnarray}
\tilde{Kl}(n,m,c,r)&=&\sum_{s|(c,r)}\sum_{\substack{0\leq a',d'<c'\\ a'd'=1\text{ mod}(c')}}\sum_{b''=0}^{s-1}e^{2\pi in\frac{d'}{c'}+2\pi im \frac{r}{s^2} \frac{a'}{c'}+2\pi i m\frac{b''}{s}}\\
&=&\sum_{s|(c,r,m)}\sum_{\substack{0\leq a',d'<c'\\ a'd'=1\text{ mod}(c')}}s\,e^{2\pi in\frac{d'}{c'}+2\pi im \frac{r}{s^2} \frac{a'}{c'}}\\
&=&\sum_{s|(c,r,m)}s\,Kl(n,mr/s^2,c/s)\label{modified Selberg id 1},
\end{eqnarray}
where we have used the fact that summing over $b''$ imposes the condition $s|m$. The modified Kloosterman sum does not depend on which decomposition, "left" or "right", we choose, so we must also have
\begin{equation}
\tilde{Kl}(n,m,c,r)=\sum_{s|(c,r,n)}s\,Kl(nr/s^2,m,c/s)\label{modified Selberg id 2},
\end{equation}
after using a "right" decomposition. Equality of both (\ref{modified Selberg id 1}) and (\ref{modified Selberg id 2}) is precisely the equality (\ref{main Kloos Selberg formula}), with $m$ in (\ref{main Kloos Selberg formula}) playing the role of $r$ in the modified Kloosterman sum.

We may wonder if the action of multiple Hecke operators  on modular forms leads to further arithmetic properties of the classical Kloosterman sums. The action of the Hecke operators $T_r$ and $T_{r'}$ on a modular form is \cite{Zagier1992}
\begin{equation}
T_rT_{r'}\circ \phi(\tau)=\sum_{d|(r,r')}d^{\omega-1}T_{rr'/d^2}\circ \phi(\tau).
\end{equation}
From this we deduce that the Fourier coefficients $\tilde{c}_{rr'}(n)$ of $\tilde{\phi}(\tau)=T_rT_{r'}\circ \phi(\tau)$ are given by
\begin{equation}
\tilde{c}_{rr'}(n)=\sum_{d|(r,r')}d^{\omega-1}\sum_{s|(n,rr'/d^2)}s^{\omega-1}c(nrr'/d^2s^2)=\sum_{\substack{s=ab\\ a|(r,r')\\ b|(n,rr'/a^2)}}s^{\omega-1}N(s)c(nrr'/s^2),
\end{equation}
where $N(s)$ is the number of ways of writing $s=ab$. Given this we can determine the map between the polar terms of $\phi(\tau)$ and $\tilde{\phi}(\tau)$ as we did before. We find
\begin{eqnarray}
&&\frac{\tilde{n}_prr'}{s^2d^2}=n_p,\;s|(\tilde{n}_p,rr'/d^2),\;d|(r,r')\Leftrightarrow\\
&&\tilde{n}_p=\frac{n_p rr'}{d^2s'^2},\;s'|(n_p,rr'/d^2),\;d|(r,r'),\;s'=\frac{rr'}{d^2s}.
\end{eqnarray}
Now we write the Rademacher expressions for both $c_{rr'}(n)$ and $c(n)$, and use the divisor sum rule to exchange the different sums over $d,s'$ and $c$, as we did in the other examples. After some algebra we find
\begin{equation}\label{arithmetic prop multi Hecke}
\sum_{\substack{s=ab\\ a|(r,r',c)\\b|(n_p,rr'/a^2,c/a)}}sN(s)Kl(n,n_prr'/s^2,c/s)=\sum_{\substack{s=ab\\ a|(r,r',c)\\b|(n,rr'/a^2,c/a)}}sN(s)Kl(nrr'/s^2,n_p,c/s).
\end{equation}
When $(r,r')=1$  we have $N(s)=1$ and we recover the expression (\ref{main Kloos Selberg formula}) as expected. In terms of the modified Kloosterman sums, the left hand-side expression can be written as
\begin{eqnarray}
\sum_{\substack{s=ab\\ a|(r,r',c)\\b|(n_p,rr'/a^2,c/a)}}sN(s)Kl(n,n_prr'/s^2,c/s)=\sum_{d|(r,r',c)}d\,\tilde{Kl}(n,n_p,c/d,rr'/d^2).
\end{eqnarray}
In fact, this is a different way of showing the equality (\ref{arithmetic prop multi Hecke}) if we use the identities for the modified Kloosterman sums. One decomposition gives the LHS and the other decomposition gives the RHS of (\ref{arithmetic prop multi Hecke}). Later on, we will use a similar property to derive the non-primitive formula for $1/8$-BPS dyons with $\psi(q)>1$ (\ref{primitivity cond 1}).

\subsection{Generalized Kloosterman Sums}\label{sec Gen Kloos sums}

In this section we apply the  logic used previously to generate arithmetic properties of generalized Kloosterman sums. These are the analog of the classic Kloosterman sums in the case of the Rademacher expansion of the Fourier coefficients of Jacobi forms. As we explain shortly, the essential difference between the classical and the generalized sums lies on the multiplier systems, which arises due to the Jacobi nature of the modular object.

 Since we will be considering the Fourier coefficients of Jacobi forms, we will need two ingredients for the derivation. The first is the action of the Hecke operators on the Jacobi forms, and for this we follow closely \cite{Eichler:1985ja}. The second is the generalized Rademacher expansion of the Fourier coefficients, which we can borrow from \cite{Dijkgraaf:2000fq,Manschot:2007ha}. 
 
 Consider a Jacobi form $\phi_{\omega,k}(\tau,z)$ of weight $\omega$ non-positive and index $k$. We take the weight to be non-positive so we can use the Rademacher expansion. The Fourier coefficients are defined from 
 \begin{equation}
 \phi_{\omega,k}(\tau,z)=\sum_{n,l} c_k(n,l)q^ny^l,\;\;q=e^{2\pi i \tau},\;\;y=e^{2\pi i z}.
 \end{equation}
 The Fourier coefficients that have $n-l^2/4k>0$ admit  the generalized Rademacher expansion \cite{Dijkgraaf:2000fq,Manschot:2007ha}
 \begin{equation}
 c_k(n,l)=\sum_{\substack{
 	n_p,l_p\\ n_p-l_p^2/4k<0}}c_k(n_p,l_p)\sum_{c=1}^{\infty}\frac{1}{c}Kl(n,l;n_p,l_p;k,c)\int_{\epsilon -i\infty}^{\epsilon+i\infty}\frac{dt}{t^{5/2-\omega}}\exp{\left[2\pi\frac{\Delta}{ct}-2\pi(n_p-l_p^2/4k)\frac{t}{c}\right]}.
 \end{equation}
 Here the sum over $n_p,l_p$ is a sum over the polar terms which have negative discriminant $n_p-l_p^2/4k<0$. The function $Kl(n,l;n_p,l_p;k,c)$ is the generalized Kloosterman sum of index $k$.

 We can now repeat the exercise of the previous section for Jacobi forms. The Hecke operator $V_r$ takes a Jacobi form of index $k$ to a Jacobi form of same weight but index $kr$. So our starting point is the formula for the Fourier coefficients of $\tilde{\phi}_{\omega,kr}(\tau,z)=V_r\circ \phi_{\omega,k}(\tau,z)$, with $\phi_{\omega,k}(\tau,z)$, a Jacobi form of weight $\omega\leq 0$ and index $k$. Following \cite{Eichler:1985ja}, the Fourier coefficients $\tilde{c}_{kr}(n,l)$ of $\tilde{\phi}_{\omega, kr}(\tau,z)$ are related to the Fourier coefficients of $\phi_{\omega,k}$ in the following way
\begin{equation}\label{Hecke sum of divisors}
\tilde{c}_{kr}(n,l)=\sum_{s|(n,l,r)}s^{\omega-1}c_{k}(nr/s^2,l/s).
\end{equation}
From this we can determine the map between the polar terms of $\tilde{\phi}_{\omega,k}$  and $\phi_{\omega,k}$ as follows
\begin{eqnarray}\label{map polar states Jacobi}
&&\tilde{n}_pr/s^2=n_p,\,\tilde{l}_p/s=l_p,\qquad s|(\tilde{n}_p,r,\tilde{l}_p),\,4kn_p-l_p^2<0 \nonumber\\
\Leftrightarrow&&\tilde{n}_p=n_pr/d^2,\,\tilde{l}_p=l_pr/d,\qquad d|(n_p,r),
\end{eqnarray} 
where $n_p,l_p$ are associated with the polar terms of $\phi_{\omega,k}$ and $\tilde{n}_p,\tilde{l}_p$ are associated with the polar terms of $\tilde{\phi}_{\omega,kr}$. The analysis of the spectrum of polar terms is quite more involved in this case as compared with the case of modular forms, described in the previous section. The reason is that, while for the modular form we have only a finite number of polar terms, for the Jacobi form there is an infinite number of polar terms $(n_p,l_p)$ due to the elliptic symmetry of the Jacobi form.  Under this transformation, also known as spectral flow transformation, a polar term with $(n_p,l_p)$ goes to
\begin{equation}\label{spectral flow transf}
n_p\rightarrow n_p(m)= n_p+l_pm+km^2,\;l_p\rightarrow l_p(m)=l_p+2km,\;m\in \mathbb{Z}.
\end{equation}
This transformation leaves the discriminant $n_p-l_p^2/4k<0$ invariant.
To find the polar terms we need to understand which points $(n_p(m),l_p(m))=(n_p+l_pm+km^2, l_p+2km)$ in the spectral flow orbit obey the conditions (\ref{map polar states Jacobi}).  Note that a spectral flow transformation in $(n_p,l_p)$ does not lead necessarily to spectral flow equivalent $(\tilde{n}_p,\tilde{l}_p)$. For example, take $d|r$ and write $m=m'\text{ mod}(d)$, with $0\leq m'<d$. From the map (\ref{map polar states Jacobi}), we have
\begin{equation}\label{spectral flow of tilde l_p}
\tilde{l}_p\rightarrow \tilde{l}_p+2km'\frac{r}{d}\text{ mod}(2kr),\;d|r,
\end{equation}
when $l_p\rightarrow l_p+2km$. So the image of $\tilde{l}_p$ under the map (\ref{spectral flow transf}) is spectral flow equivalent to $\tilde{l}_p$ when $m'=0$. This is important because in the spectral flow orbit of $(n_p,l_p)$ there can be some $n_p$ with the property that $d|n_p$, for $d|r$, but associated to spectral flow inequivalent $\tilde{l}_p$. To be more explicit, take $m=m'\text{ mod}(d)$ with $0\leq m'<d$, then we have $n_p+l_pm+km^2=n_p+l_pm'+km'^2\text{ mod}(d)$. Suppose there are $N$ different $m'$ for which $d|(n_p+l_pm'+km'^2)$. Because of the transformation (\ref{spectral flow of tilde l_p}), this gives rise to $N$ inequivalent $(\tilde{n}_p,\tilde{l}_p)$ terms.

We proceed as in the previous section and write the Rademacher expansion of $\tilde{c}_{kr}(n,l)$. Using the map between the polar coefficients we obtain
\begin{eqnarray}
\tilde{c}_{kr}(n,l)=&&\sum_{\substack{n_p,l_p\\ 4kn_p-l_p^2<0}}c_{k}(n_p,l_p)\sum_{\substack{d|r\\ 0\leq m<d\\m:\, d|n_p(m)}}\left(\frac{r}{d}\right)^{\omega-1}\sum_{c=1}^{\infty}\frac{1}{c}Kl\left(n,l;n_p(m)r/d^2,l_p(m)r/d;kr,c\right) \nonumber\\
&&\times\int_{\epsilon-i\infty}^{\epsilon+i\infty} \frac{dt}{t^{5/2-\omega}}\exp{\left[2\pi\frac{\Delta}{ct}- 2\pi\frac{r}{d^2}(n_p-l_p^2/4k)\frac{t}{c} \right]}.
\end{eqnarray}
We have used the fact that $n_p(m)-l_p(m)^2/4k=n_p-l_p^2/4k$. The sum over $m$ is due to the sum over spectral flow inequivalent $(\tilde{n}_p,\tilde{l}_p)$ terms. We rescale $t$ as $t\rightarrow td/r$ to obtain
\begin{eqnarray}
\tilde{c}_{kr}(n,l)=&&\sum_{\substack{n_p,l_p\\ 4kn_p-l_p^2<0}}c_{k}(n_p,l_p)\sum_{\substack{d|r\\ 0\leq m<d\\m:\, d|n_p(m)}}\left(\frac{r}{d}\right)^{1/2}\sum_{c=1}^{\infty}\frac{1}{c}Kl\left(n,l;n_p(m)r/d^2,l_p(m)r/d;kr,c\right) \nonumber\\
&&\times\int_{\epsilon-i\infty}^{\epsilon+i\infty} \frac{dt}{t^{5/2-\omega}}\exp{\left[\frac{\Delta r}{cdt}- (n_p-l_p^2/4k)\frac{t}{cd} \right]}.
\end{eqnarray}
Now we use the divisor sum rule (\ref{sum c sum s}) to exchange the sums over $d$ and $c$. We obtain
\begin{eqnarray}\label{Rademacher Jacobi 1}
\tilde{c}_{kr}(n,l)=&&\sum_{\substack{n_p,l_p\\ 4kn_p-l_p^2<0}}c_{k}(n_p,l_p)\sum_{c=1}^{\infty}\frac{1}{c}\sum_{\substack{d|(r,c)\\ 0\leq m<d\\m:\, d|n_p(m)}}{(rd)}^{1/2}Kl(n,l;n_p(m)r/d^2,l_p(m)r/d;kr,c/d)\nonumber\\
&&\times\int_{\epsilon-i\infty}^{\epsilon+i\infty} \frac{dt}{t^{5/2-\omega}}\exp{\left[2\pi\frac{\Delta r}{ct}-2\pi (n_p-l_p^2/4k)\frac{t}{c} \right]}.
\end{eqnarray}

Since $\tilde{c}_{kr}(n,l)$ can be written as the sum  (\ref{Hecke sum of divisors}), plugging back the Rademacher expansion of $c_k(n,l)$ in the formula (\ref{Hecke sum of divisors}) and comparing with the expansion (\ref{Rademacher Jacobi 1}) leads to non-trivial arithmetic properties of the Kloosterman sums. In particular, we must have the following equality 
\begin{eqnarray}\label{Selberg alpha id}
&&\sum_{d|(r,c)}\sum_{\substack{0\leq m<d\\m:\, d|n_p(m)}}(rd)^{1/2}Kl(n,l;n_p(m)r/d^2,l_p(m)r/d;kr,c/d)=\nonumber\\
&&\qquad=\sum_{s|(n,l,r,c)} s^{3/2}\,Kl(nr/s^2,l/s;n_p,l_p;k,c/s).
\end{eqnarray}
To show that this is true, we plug (\ref{Selberg alpha id}) back in (\ref{Rademacher Jacobi 1}) and rescale $t$ as $t\rightarrow ts$,
\begin{eqnarray}\label{Rademacher Jacobi 2}
\tilde{c}_{kr}(n,l)=&&\sum_{\substack{n_p,l_p\\ 4kn_p-l_p^2<0}}c_{k}(n_p,l_p)\sum_{c=1}^{\infty}\frac{1}{c} \sum_{s|(n,l,r,c)} s^{\omega}\,Kl(nr/s^2,l/s;n_p,l_p;k,c/s)\nonumber\\
&&\times\int_{\epsilon-i\infty}^{\epsilon+i\infty} \frac{dt}{t^{5/2-\omega}}\exp{\left[2\pi\frac{\Delta r/s^2}{t(c/s)}-2\pi (n_p-l_p^2/4k)\frac{t}{c/s} \right]}.
\end{eqnarray}
Exchanging again the sums over $c$ and $s$ using the formula (\ref{sum c sum s}) we obtain
\begin{eqnarray}
\tilde{c}_{kr}(n,l)=&&\sum_{s|(n,l,r)} s^{\omega-1}\sum_{\substack{n_p,l_p\\ 4kn_p-l_p^2<0}}c_{k}(n_p,l_p)\sum_{c=1}^{\infty}\frac{1}{c}\,Kl(nr/s^2,l/s;n_p,l_p;k,c) \nonumber\\
&&\times\int_{\epsilon-i\infty}^{\epsilon+i\infty} \frac{dt}{t^{5/2-\omega}}\exp{\left[2\pi\frac{\Delta r/s^2}{tc}-2\pi (n_p-l_p^2/4k)\frac{t}{c} \right]}\nonumber\\
=&&\sum_{s|(n,l,r)} s^{\omega-1}c_k(nr/s^2,l/s),
\end{eqnarray}
where we have used the Rademacher expansion of $c_k(n,l)$ to obtain the second line. This demonstrates that we have indeed (\ref{Selberg alpha id}), as we wanted to show.

In the following, we use a different method to show the arithmetic equality (\ref{Selberg alpha id}). As in the example of the classical Kloosterman sums, we construct modified versions of the generalized Kloosterman sums. In \cite{Gomes:2017bpi},  an analytic formula for the generalized Kloosterman sums was developed. We will use it here extensively. The exercise is very similar to the classical Kloosterman case and we build new sums based on two by two matrices with determinant $r$. The arithmetic equality between the generalized Kloosterman sums then follows from different choices of decomposing that matrix, much like we did in the previous section.   Moreover, the modified version will play a central role in the holographic computation, which is why it is important to develop its properties here. 

We define the modified versions as
\begin{equation}\label{modified gen Kloos}
	\tilde{Kl}(n,l;n_p,l_p;k,c,r)=\sum_{\substack{0\leq a,d<c\\ ad=r\text{ mod}(c)}}e^{2\pi i (n-l^2/4k) \frac{d}{c}+2\pi i(n_p-l_p^2/4k)\frac{a}{c} } M(\gamma)_{l, l_p},\quad \gamma=\left(\begin{array}{cc}
	a & b\\
	c & d
	\end{array}\right),\,\text{det}(\gamma)=r,
\end{equation}
where $M(\gamma)_{l,l_p}$ is a multiplier matrix. This matrix has the following analytic expression
\begin{equation}
M(\gamma)_{l,l_p}=\frac{1}{\sqrt{2i k c}}\sum_{m=0}^{c-1}\exp{\left[\frac{\pi i}{2k}\frac{a}{c}(l_p+2k m)^2-\frac{\pi i}{k c}(l_p+2k m)l+\frac{\pi i}{2k}\frac{d}{c}l^2\right]}.
\end{equation}
For $r=1$ we recover the definition of the generalized Kloosterman sums described in \cite{Gomes:2017bpi}. Note that this expression is explicitly invariant under $l_p\rightarrow l_p \text{ mod}(2k)$ \footnote{The shift of $l_p$ by $\text{mod}(2k)$ can be compensated by a shift in $n\in \mathbb{Z}/c\mathbb{Z}$, which is being summed over. Further details can be found in \cite{Gomes:2017bpi}.}. However, it is more involved to show that it is also invariant under $l \rightarrow l \text{ mod}(2kr)$. In the following, we confirm that it is the case by showing that the modified Kloosterman sums can be written as a sum over generalized Kloosterman sums of index $kr$. 

Lets first consider a "right" decomposition of the matrix $\gamma$, which preserves the ratios $a/c=a'/c'$. That is, we write
\begin{equation}
\left(\begin{array}{cc}
a & b\\
c & d
\end{array}\right)=\left(\begin{array}{cc}
a' & b'\\
c' & d'
\end{array}\right)\left(\begin{array}{cc}
s & b''\\
0 & r/s
\end{array}\right),\;\left(\begin{array}{cc}
a' & b'\\
c' & d'
\end{array}\right)\in PSL(2,\mathbb{Z}).
\end{equation}
This means we have
\begin{equation}
c=c's,\;\frac{a}{c}=\frac{a'}{c'},\;\frac{d}{c}=\frac{b''}{s}+\frac{d'}{c'}\frac{r}{s^2},
\end{equation}
with $0 \leq a',d' <c'$ and $0\leq b''<s$. Note that the $d/c$ dependence in the multiplier matrix cancels against a similar term in the first exponential of the expression (\ref{modified gen Kloos}). A right decomposition leads to the following expression 
\begin{eqnarray}
	\tilde{Kl}(n,l;n_p,l_p;k,c,r)=&&\sum_{s|(c,r)}\sum_{\substack{0\leq a',d'<c'\\ a'd'=1\text{ mod}(c')}}\sum_{b''=0}^{s-1}e^{2\pi in\frac{b''}{s}}e^{2\pi i n\frac{r}{s^2} \frac{d'}{c'}-2\pi i(n_p-l_p^2/4k)\frac{a'}{c'} }\nonumber\\
	&&\times\frac{1}{\sqrt{2i k c's}}\sum_{m=0}^{c's-1}\exp{\left[\frac{\pi i}{2k}\frac{a'}{c'}(\mu+2k m)^2-\frac{\pi i}{k c's}(\mu+2k m)l\right]}.\label{sum over n}
\end{eqnarray}
Further, we write $m=m'+jc'$ with $0\leq m'<c'$ and $0\leq j<s$. The sum over $m$ in (\ref{sum over n}) becomes the double sum
\begin{equation}
\sum_{m'=0}^{c'-1}\sum_{j=0}^{s-1}\exp{\left[-2\pi i\frac{lj}{s}+\frac{\pi i}{2k}\frac{a'}{c'}(\mu+2k m')^2-\frac{\pi i}{k c's}(\mu+2k m')l\right]}.
\end{equation}
The sum over $j$ imposes that $s|l$. Similarly, the sum over $b''$ imposes $s|n$. From both the sums over $j,b''$ we obtain a factor of $s^2$. From the $1/\sqrt{c}$ normalization of the multiplier matrix there is an additional $1/s^{1/2}$ factor, so in total there is a $s^{3/2}$ multiplicative factor. We can recast the final result as a sum over generalized Kloosterman sums, that is,
\begin{equation}\label{modified Selberg id gen Kloos 1}
	\tilde{Kl}(n,l;n_p,l_p;k,c,r)=\sum_{s|(c,r,n,l)}s^{3/2} Kl(nr/s^2,l/s;n_p,l_p;k,c/s).
\end{equation}

Using instead a "left" decomposition, we have 
\begin{equation}
c=c's,\;\frac{d}{c}=\frac{d'}{c'},\;\frac{a}{c}=\frac{b''}{s}+\frac{a'}{c'}\frac{r}{s^2},
\end{equation}
with $0 \leq a',d' <c'$ and $0\leq b''<s$.  The part of the modified Kloosterman sum that is relevant in this computation is
\begin{eqnarray}
&&e^{2\pi i (n_p-l_p^2/4k)\frac{a}{c}}\sum_{m=0}^{c-1}\exp\left[\frac{\pi i}{2k}\frac{a}{c}(l_p+2k m)^2-\frac{\pi i}{kc}(l_p+2k m)l
\right]\nonumber\\
&&=\sum_{m=0}^{c-1}\exp{\left[2\pi i (n_p+ml_p+km^2)\frac{b''}{s}\right]}\exp{\left[2\pi i\frac{a'}{c'}\frac{r}{s^2}(n_p+l_pm+km^2)-\frac{\pi i}{kc}(l_p+2k m)l\right]}.\nonumber\\
{}\label{left decomp gen Kloos}
\end{eqnarray}
After writing $m=m'+js$ with $0\leq m'<s$ and $0\leq j<c'$, the sum over $b''$ becomes
\begin{equation}
\sum_{b''=0}^{s-1}e^{2\pi i(n_p+m'l_p+km'^2)\frac{b''}{s}}=s \delta(s|n_p+m'l_p+km'^2)=s\delta(s|n_p(m')),
\end{equation}
 where $\delta$ is the Kroneker delta function. The expression (\ref{left decomp gen Kloos}) becomes
 \begin{eqnarray}
 \sum_{\substack{0\leq m'<s\\ m': \,s|n_p(m')}}s\, e^{2\pi i \left(\frac{n_p(m')r}{s^2}-\frac{(l_p (m')r/s)^2}{4kr}\right)\frac{a'}{c'}}\sum_{j=0}^{c'-1}\exp\left[\frac{\pi i}{2kr}\frac{a'}{c'}(l_p(m')r/s +2kr j )^2-\frac{\pi i}{k r c'}(l_p(m')r/s+ 2k r j)l
 \right].\nonumber \\
 {}\label{left decomp gen Kloos 2}
 \end{eqnarray}
From the overall normalization of the modified Kloosterman sum $1/\sqrt{c}=1/(\sqrt{c'}\sqrt{s})$ we obtain a factor of $1/s^{1/2}$. Together with the factor of $s$ in (\ref{left decomp gen Kloos 2}) we obtain an overall factor of $s^{1/2}$. Putting back the $d/c$ dependence of the Kloosterman sum in (\ref{left decomp gen Kloos 2}) we find the following sum over generalized Kloosterman sums of index $kr$
\begin{equation}\label{modified Selberg id gen Kloos 2}
\tilde{Kl}(n,l;n_p,l_p;k,c,r)=\sum_{s|(r,c)}\sum_{\substack{0\leq m<s \\ m:\,s|n_p(m)}}(rs)^{1/2}Kl(n,l;n_p(m)r/s^2,l_p(m)r/s;kr,c/s).
\end{equation}
Equality of (\ref{modified Selberg id gen Kloos 1}) and (\ref{modified Selberg id gen Kloos 2}) gives back the identity (\ref{Selberg alpha id}).

\section{Holographic Computation}\label{sec hol comp}

In this section, we consider the holographic computation of the black hole degeneracy. We consider the path integral on the $AdS_2\times S^1\times S^2/\mathbb{Z}_c$ orbifold geometries with at most $\mathbb{Z}_r$ conical singularities.  Using the results of \cite{Gomes:2017eac,Gomes:2017bpi,Dabholkar:2014ema}, we find that the partition function on each geometry has the form of a Bessel function multiplied by a modified generalized Kloosterman sum.  Summing over arbitrary $\mathbb{Z}_c$ orbifolds, we reproduce the structure of the non-primitive degeneracy formulas (\ref{non-primitive deg}), including the dependence on the U-duality invariants. The focus of this section will be on the Bessel function and the Kloosterman sums. Later we apply the results of this section to the $\CN=8$ and $\CN=4$ compactifications and compare with the microscopic degeneracies. As explained in \cite{Gomes:2017eac}, what effectively distinguishes the compactifications is the spectrum of polar states. In both the $\CN=8$ and $\CN=4$ examples the data about the internal manifold, transverse to the $AdS_2\times S^1\times S^2/\mathbb{Z}_c$ geometry,  remains unchanged when $r\neq1$. This is the reason why the approach followed in this section can be systematically applied in both compactifications.

The $AdS_2\times S^1\times S^2/\mathbb{Z}_c$ orbifold consists of a $2\pi/c$ identification along the angles of the $AdS_2$ (Euclidean) and the sphere $S^2$, together with an identification on the circle $S^1$ of the form $y\sim y+2\pi R d/c$, with $R$ the radius.  These type of orbifolds were described in more detail in \cite{Sen:2009vz,Murthy:2009dq,Dabholkar:2014ema,Gomes:2017bpi}. However, in those works the orbifold is always smooth by requiring the integers $(c,d)$ to be co-prime. Here we relax that condition. The $AdS_2\times S^1/\mathbb{Z}_c$ factor of the orbifold that  we denote as $M_{(c,d)}$, is topologically a solid torus, that is, $M_{(c,d)}\simeq D\times S^1$, with $D$ a disk. Globally, the orbifold geometry changes which cycles of the boundary torus become contractible and non-contractible in the full geometry \cite{Dijkgraaf:2000fq}. For each geometry $M_{(c,d)}$ we bound a disk $D$ to a cycle $C_c$ on the boundary torus. The cycle $C_c$ is thus contractible in the full geometry. On the other hand, the circle parametrizes the cycle $C_{nc}$, which is thus non-contractible. In a basis $C_1$  and $C_2$ of one-cycles of the boundary torus, with intersection $C_1\cap C_2=1$, the filling of the $M_{(c,d)}$ geometry is such that
\begin{equation}\label{cycles map}
C_{nc}=aC_1+bC_2,\;C_c=cC_1+dC_2,
\end{equation}
with 
\begin{equation}
ad-bc=r,\quad a,d,c,b\in\mathbb{Z}.
\end{equation}
As an example, take $c=0$, which implies $d|r$. So if we take $d=r$ the contractible cycle is $C_c=rC_2$ which leads to a $\mathbb{Z}_r$ singularity at the origin of the disk. In general, the orbifold will lead to singularities whenever $\text{gcd}(c,d)\neq 1$. The orbifold has a  $\mathbb{Z}_s$ conical singularity whenever $s=\text{gcd}(c,d)$ for $s|r$. 

The $\mathbb{Z}_r$ singular orbifolds arise after a M-theory uplift of $r\D6-r\aD6$ configurations with fluxes. The uplift to M-theory consists of a pair of $r \text{TN}-r\bar{\text{TN}}$ with $G$-flux turned on along the Calabi-Yau directions ($T^6$ or $K3\times T^2$ in the present work). A decoupling limit from the asymptotic region gives back the orbifold geometry \cite{deBoer:2008fk}. The proposal of \cite{Gomes:2017eac} considers a generalization of these geometries to include "singular" $G$-fluxes. On the $\D6$-brane worldvolume theory we can consider $U(1)$ field strengths that have a singular component but with well defined Chern-classes. The singular fluxes can be regularized using the notion of Ideal sheaves \cite{Iqbal:2003ds}. The crucial aspect in this construction \cite{Gomes:2017eac} is that the M-theory uplift of the $\D6$-brane with singular fluxes, leads to non-trivial $G$-fluxes along the Calabi-Yau directions. Furthermore, one finds that the Kahler form of the Calabi-Yau is corrected by a term proportional to this flux, precisely as in the quantum foam and Kahler gravity picture described in \cite{Iqbal:2003ds}. 

The sum over the singular $G$-fluxes indicates that the geometry of the Calabi-Yau is fluctuating wildly as explained in \cite{Iqbal:2003ds}. In the K\"{a}hler gravity picture we are summing over different geometries in the path integral. The work of \cite{Gomes:2017eac} attempts to provide with a low energy effective action for the theory on these K\"{a}hler geometries. It is argued that the presence of such G-flux leads to a finite renormalization of the parameters that define the five dimensional Lagrangian. In particular, it is argued that for the $T^6$  and $K3\times T^2$ compactifications, the coefficient that parametrizes a mixed gauge-gravitational Chern-Simons term, which in the theory without singular fluxes is proportional to the second Chern-class of the tangent bundle of the Calabi-Yau, is shifted by the second Chern-class of the singular $U(1)$ fluxes. Another important aspect of this construction, which arises from the fact that we can consider different $U(1)$-bundles on the $\D6$ and $\aD6$-branes, is that at the on-shell level the five dimensional abelian gauge fields acquire non-trivial holonomies along the $AdS_2$ disk. This comes from the fact that such flux configurations have an equivalent description in terms of $\M2$ and $\aM2$-branes wrapping two-cycles of the Calabi-Yau and sitting at the origin of the disk \cite{Gaiotto:2006ns,Denef:2007vg}; the corresponding Wilson lines are proportional to the total $\M2$ charge.  

With the aforementioned renormalizations and the particular abelian gauge fields, \cite{Gomes:2017eac} computes the five dimensional path integral on the $AdS_2\times S^1\times S^2$ background using supersymmetric localization. For each geometry, one obtains a Bessel function with a non-trivial dependence on the Chern-classes of the singular $G$-fluxes. This dependence  generates the polarity that we observe in the Rademacher expansion. In particular, the inclusion of the $\M2/\aM2$ abelian gauge field configurations is responsible for the spectral flow sector dependence that one finds at the level of the Bessel function \cite{Gomes:2017eac,Gomes:2017bpi}. 

The sum over the fluxes is not arbitrary, nevertheless. Due to the renormalizations, the physical size of the $AdS_2\times S^1\times S^2$ background gets quantum corrected and can become zero if we include sufficiently large Chern-classes. In the $\D6,\aD6$ picture the size of the metric can be related to the distance between the $\D$-branes, and so the geometry ceases to exist when the $\D$-branes collapse on top of each other. Therefore, due to this condition on the physical size, we have only a finite number of these geometries. The bound on this number has been related to the stringy exclusion principle \cite{Gomes:2017eac}.

Another important result that follows from the proposal \cite{Gomes:2017eac} is the explanation of an exact formula for the generalized Kloosterman sums and corresponding Bessel functions that appear in the counting of primitive $\CN=4$ dyons \cite{Gomes:2017bpi}. The microscopic counting formulas are mock-modular Jacobi forms \cite{Dabholkar:2012nd}, but albeit their unusual modular properties, the Fourier coefficients have similar Rademacher expansions \cite{Ferrari:2017msn} as in the Jacobi-form case. The work of \cite{Gomes:2017bpi}  generalizes the construction of \cite{Dabholkar:2014ema} to Kloosterman sums of Jacobi forms of arbitrary index. The generalized Kloosterman sums are explained by a sum over flat connections in Chern-Simons theory with microcanonical boundary conditions. The Chern-Simons action of the flat connections reproduces the various phases that are characteristic of the Kloosterman sums. After taking into account the holonomies induced by the fluxes, whose Wilson lines are proportional to the total $\M2$ charge in the equivalent picture, the generalized Kloosterman sums acquire the dependence on the spectral flow sectors characterized by the integers $l,l_p$ in the multiplier matrices  \cite{Gomes:2017bpi}. However, in the computation of \cite{Gomes:2017bpi} only single $\D6-\aD6$ configurations were considered. In this case, the geometry $M_{(c,d)}$ has $\text{gcd}(c,d)=1$ and the gravitational computation reproduces the generalized Kloosterman sums that appear in the Rademacher expansion. 

In this section, we review the computation of the Kloosterman sums and extend it to the orbifold geometries with Dhen filling $ad-bc=r$ following \cite{Dabholkar:2014ema,Gomes:2017bpi}.

\subsection{Modified Kloosterman Sums from Chern-Simons Theory}\label{sec mod Kloos CS}

Our task is to reproduce the results for the $AdS_2$ partition function derived in \cite{Gomes:2017bpi}, but on the $AdS_2\times S^1\times S^2/\mathbb{Z}_c$ orbifolds with at most $\mathbb{Z}_r$ conical singularities. As explained  in the beginning of section  \S\ref{sec hol comp}, this corresponds to geometries $M_{(c,d)}$ with filling parametrized by the integers $ad-bc=r$. The computation in these orbifolds will give us directly the modified Kloosterman sums multiplied by the corresponding Bessel function. 

 There are two main contributions to the partition function on the $AdS_2\times S^1\times S^2/\mathbb{Z}_c$ orbifold. The first comes from a set of continuous modes, which parametrize fluctuations around the on-shell background. These modes are determined by the localization off-shell solutions computed on $AdS_2\times S^1\times S^2$  \cite{Gomes:2013cca,Gomes:2017eac} which are a five dimensional uplift of the solutions originally derived in \cite{Dabholkar:2010uh,Gupta:2012cy}. From a four dimensional point of view, these solutions correspond to normalizable modes of the various vector-multiplet scalar fields, while from a five dimensional point of view they correspond to normalizable fluctuations of the radius of the circle  $S^1$ and the Wilson lines of the abelian gauge fields along the circle $S^1$. Integration over these modes in the unorbifolded geometry, results in the Bessel function \cite{Gomes:2017eac}
 \begin{equation}\label{pt fnct}
 Z_{AdS_2\times S^1\times S^2}(f,\bar{f})=\int_{\epsilon -i\infty}^{\epsilon +i\infty} \frac{dt}{t^{1+b_2/2}}\exp{\left[2\pi \frac{\hat{q}_0}{t}+2\pi \Delta_{\text{polar}}(f,\bar{f})t\right]},
 \end{equation}
 with $\hat{q}_0$  the quadratic charge invariant
 \begin{equation}
 \hat{q}_0=-q_0+D^{ab}q_aq_b/2,
 \end{equation}
 and $\Delta_{\text{polar}}(f,\bar{f})$ is the polarity, which depends explicitly on the singular fluxes $f,\bar{f}$ by means of their integer Chern-classes.  Here, $D_{ab}=D_{abc}p^c$ and $D_{abc}$ is the intersection matrix of the Calabi-Yau. The integers $p^a$ parametrize magnetic flux on the sphere $S^2$. We furthermore enforce $p^1=1$ to make contact with the analysis of \cite{Gomes:2017bpi}. For the compactifications on $K3\times T^2$ and $T^4\times T^2$ we have $D_{abc}\equiv D_{ij1}=D_{1ij}=D_{i1j}=C_{ij}$ with $i=2\ldots \text{dim }H^{2}(K3,T^4)+1$.  In terms of the charges $q_a,p^a$, the  T-duality invariants are computed as follows
 \begin{equation}\label{Cqq 1}
 Q^2/2=-q_0+C^{ij}q_iq_j/2,\; P^2/2=C_{ij}p^ip^j,\; Q.P=q_1-q_ip^i.
 \end{equation}
 This gives
 \begin{equation}\label{Cqq 2}
 \hat{q}_0=\big(Q^2P^2-(Q.P)^2\big)/2P^2.
 \end{equation}
  The integration variable $t$ can be identified with the radius of the circle $S^1$ evaluated at the origin of $AdS_2$.  The index of the Bessel function depends on which compactification we are considering. In \cite{Gomes:2015xcf}, the values for the index were computed 
 \begin{equation}
 \CN=8:\;b_2=7,\qquad \CN=4:\; b_2=k+1,
 \end{equation}
 where $k$ is the number of $\CN=4$ vector-multiplet fields in the four dimensional CHL compactification . For example, in the case of $K3\times T^2$ we have $k=22$. For the CHL compactifications we can identify $b_2$ with the dimension of the  $H^{2}(K3\times T^2/\mathbb{Z}_N)$, where $\mathbb{Z}_N$ is the CHL orbifold group. In the $\CN=8$ case, one has to consider a $\CN=4$ subalgebra of the $\CN=8$ supersymmetry algebra \cite{Sen:2008sp}. In this truncation, $b_2=7$ is the number of $\CN=4$ vector-multiplet fields.
 
 The path integral on the smooth $\mathbb{Z}_c$ orbifolds, with $\text{gcd}(c,d)=1$, reproduces  the subleading Bessel functions parametrized by $c$ in the Rademacher expansion \cite{Gomes:2017bpi}. For the case of the $\mathbb{Z}_c$ orbifolds with conical singularities we are instructed to perform two main tasks. The first is to rederive the Bessel  function (\ref{pt fnct}) for the $M_{(c,d)}$ geometries with $ad-bc=r$. This includes determining the precise dependence of the Bessel function on the parameter $r$ and understand the exact spectrum of the polar states and their polarity $\Delta_{\text{polar}}$. The second task is to re- consider the problem of summing over flat connections in the path integral and compute their Chern-Simons action on the orbifold geometry. These sums will give rise to the various phases that we encounter in the Kloosterman sums. In this problem, it will be enough to use an effective Chern-Simons description in three dimensions much like in \cite{Dabholkar:2014ema,Gomes:2017bpi}.  The robustness of the Chern-Simons computation follows essentially from its topological nature. Moreover, the inclusion of the flat connections in the path integral does not affect the computation of the Bessel function, which, after all, captures only local fluctuations of the quantum fields around the background. The flat connections, on the other hand, parameterize global contributions to the path integral by means of their holonomies,   which are captured only by the Chern-Simons terms.

Before moving to the actual computation, we explain in more detail how one goes from the $\D6-\aD6$ configuration to the $AdS_2$ geometry. The M-theory uplift of the $r\D6-r\aD6$ configuration with fluxes $p^a$ per $\D6,\aD6$ is the $r$ TN-$\bar{\text{TN}}$ geometry with G-fluxes proportional to $p^a$ living on the Calabi-Yau directions. This geometry admits a decoupling region near the core of the two centers where the KK monopoles are sitting. The decoupling geometry is exactly global $AdS_3\times S^2$ orbifolded by a $\mathbb{Z}_r$ group. In addition,  the sphere comes twisted by a $SU(2)$ flat connection, which ensures that it is a geometry dual to the R-sector of the CFT  \cite{deBoer:2008fk}. After a particular thermal identification of the time direction, this geometry becomes the euclidean near-horizon geometry of the $\D0-\D4$ black hole \cite{Beasley:2006us}, with the roles of euclidean time and space interchanged- this is the $AdS_2\times S^1\times S^2$ geometry which is the focus of our work. Note that from the $\D$-brane picture the total flux is proportional to $rp^a$ because we have $r\D6$ and $r\aD6$. However, after uplift, the parameter $r$ becomes part of the geometry transverse to the Calabi-Yau and the G-flux remains proportional to $p^a$ as in the case $r=1$. Following \cite{Gomes:2017eac}, this means that all the information that enters in the partition function about the Calabi-Yau, including the attractor value of its Kahler form, remains unchanged when compared with the case $r=1$. All of this remains true even after turning on the singular fluxes $f,\bar{f}$ on the Calabi-Yau. Therefore, we find that the spectrum of polar states, which is determined by the fluctuations of the fields living on the Calabi-Yau, is left unchanged and only depends on the information of the Calabi-Yau in the case $r=1$. 

 As a warm-up, we consider the theory on $AdS_2\times S^1\times S^2$, which corresponds to the orbifold $\mathbb{c}$ with $c=1$. Since we have $ad-bc=r$, with $0\leq a,d<c$, this means we must have  $a=d=0$ and $b=-r$. The map between the cycles $C_1,C_2$ on the boundary of the solid torus and the contractible and non-contractible cycles (\ref{cycles map}) becomes
\begin{equation}
C_{nc}=-rC_2,\;C_c=C_1.
\end{equation}
This means that in evaluating the physical action, we have to integrate $r$ times around the circle $S^1$, which corresponds to the cycle $C_2$. We can also see this geometry  as an $r-$cover of the $AdS_2\times S^1/\mathbb{Z}_r$ orbifold with a $\mathbb{Z}_r$ conical singularity at the origin of the $AdS_2$ disk. Following the steps described in \cite{Gomes:2017eac}, the renormalized supergravity action evaluated on the localization locus is $r$ times the renormalized action for the theory with $r=1$, that is,
\begin{eqnarray}\label{Ren action times r}
S(q,p,\phi)_r&=&rS(q,p,\phi)_1\nonumber\\
&=&\pi r\hat{q}_0\phi^0+r\frac{\pi}{6}\frac{p^3+c_2\cdot p}{\phi^0}-\frac{\pi r}{2\phi^0}D_{ab}(\phi^a+q^a\phi^0)(\phi^b+q^b\phi^0),
\end{eqnarray}
with  $S(q,p)$ the renormalized action. The zero-dimensional fields $\phi^{0,a}$ parameterize the off-shell solutions that are left unfixed under the localization procedure. In particular, they correspond to the values of those fields evaluated at the origin of $AdS_2$. In this case, we can identify $1/\phi_0$ with the radius of the circle $S^1$ for unit size of $AdS_2$. 

So far we have considered the case with no fluxes to simplify the discussion. Nevertheless, we can already see from the previous exercise that the effect of the $\mathbb{Z}_r$ orbifold in the polarity of the Bessel function, which is determined by the term that multiplies $1/\phi_0$ in (\ref{Ren action times r}) after integrating out $\phi^a$,  is to multiply the polarity of the theory with $r=1$ by $r$. In other words we have
\begin{equation}
\Delta_{\text{polar}}(p)|_{r}=r\Delta_{\text{polar}}(p)|_{r=1}.
\end{equation}
For example, in the absence of fluxes the polarity is $\Delta_{\text{polar}}=(p^3+c_2\cdot p)/24$.
Introducing the singular fluxes in the problem, as described in \cite{Gomes:2017eac}, leads to a redefinition of the parameter $c_2$ that appears in (\ref{Ren action times r}) and an additional contribution proportional to the total $\M2$-brane charge, in the equivalent picture. It is important to stress again that the information about the polarity and the spectrum of polar states is still determined by the $r=1$ theory, as explained previously. From the expression (\ref{Ren action times r}), the effective charges that multiply the chemical potentials $\phi^{0,a}$ acquire a multiplicative factor of $r$, as we can easily see from the term $ rq_0\phi^0+rq_a\phi^a$. This fact will play a very important in the following analysis.

To get acquainted with the Chern-Simons computation let us first consider the  on-shell theory both from the supergravity  and the Chern-Simons pictures. The on-shell action in the supergravity picture can be obtained by extremizing the renormalized action (\ref{Ren action times r}) with respect to $\phi^{0,a}$, that is, 
\begin{equation}\label{on-shell entropy}
S(q,p)_r|_{\text{on-shell}}=\pi r\hat{q}_0(\phi^0)^{*}+r\frac{\pi}{6}\frac{p^3+c_2\cdot p}{(\phi^0)^{*}},
\end{equation}
where $(\phi^0)^{*}$ is the on-shell value of $\phi^0$ which obeys $\partial_{\phi^0}S(q,p)_r|_{(\phi^0)^{*}}=0$. From the Chern-Simons point of view we can compare this result with the Chern-Simons action of the various flat connections  computed on the same geometry. It has been shown in \cite{Dabholkar:2014ema} that the real part of the Chern-Simons action of the flat connection  corresponds to the on-shell entropy, provided that we introduce the appropriate boundary terms consistent with the microcanonical ensemble. In this case, the value of $\phi^0$ is encoded  in the holonomy of the connection and the constant $p^3+c_2\cdot p$ is related to the Chern-Simons level. 

For the problem at hands, we have Chern-Simons terms based on the gauge group $SL(2,\mathbb{R})_L\times SL(2,\mathbb{R})_R\times SU(2)_R$ \footnote{To perform the computation in the Euclidean theory, one has to consider the complexification of the $SL(2)$ connections \cite{Dabholkar:2014ema}.}, and action
\begin{equation}\label{action w CS3}
S=-\frac{i \tilde k_L}{4\pi}I[\tilde A_L]+\frac{i \tilde k_R}{4\pi}I[\tilde A_R]-\frac{i k_R}{4\pi}I[{A}_R],
\end{equation}
with $\tilde k_L=c_L/6$ and $\tilde k_R=k_R=c_R/6$ the corresponding levels. These levels are determined by the $r=1$ theory. The fields $\tilde{A}_L$, $\tilde{A}_R$ and $A_R$ are the gauge connections of the $SL(2,\mathbb{R})_L\times SL(2,\mathbb{R})_R\times SU(2)_R$ factors respectively. The theory also contains $U(1)$ Chern-Simons but since their action vanishes for a flat connection they will not play a role at this stage. We have defined 
\begin{equation}
I[A]= \int_{D\times S^1}\text{Tr}\left(A\wedge dA+\frac{2}{3}A^3\right),
\end{equation}
as the Chern-Simons action for a particular gauge group factor.
Besides the bulk term we need to add appropriate boundary terms in order to impose microcanonical boundary conditions. These boundary terms are \cite{Dabholkar:2014ema}
\begin{equation}
I_{b}(A) = \int_{T^2}\text{Tr}A_1A_2d^2x,
\end{equation}
where $A_1$ is the component of $A$ along the cycle $C_1$ at the boundary and $A_2$ is the component  along $C_2$. The boundary conditions consist in fixing $A_2$ and letting $A_1$ to fluctuate. 

At the on-shell level, we have the flatness condition $dA+A\wedge A=0$ for the various factors in the gauge group. The action of the flat connection can be computed as in \cite{Dabholkar:2014ema,Gomes:2017bpi}, based on the original computation of \cite{kirk1990} on the solid torus.  The action for the $SL(2,\mathbb{R})_R$ and $SU(2)_R$ flat connections cancel by supersymmetry and we are left with the $SL(2,\mathbb{R})_L$ part. The bulk integral, proportional to $I[\tilde{A}_L]$, gives on the $M_{(c,d)}$ geometry 
\begin{equation}
S_{\text{bulk}}=-\pi i\frac{\tilde{k}_L}{2}\frac{a\tau+b}{c\tau+d},
\end{equation}
while the boundary term is
\begin{equation}
S_{\text{bnd}}=-i\frac{\tilde{k}_L}{4\pi}I_{b}=\frac{\pi i}{2} \tilde{k}_L r \frac{\tau}{(c\tau+d)^2}.
\end{equation}
Here $\tau$ is parametrizing the holonomy. Note that contrary to the case with $r=1$ considered in \cite{Dabholkar:2014ema,Gomes:2017bpi}, here we have to integrate the boundary term $r$ times, which gives the factor of $r$ in $S_{\text{bnd}}$. This is such that the boundary cycles preserve the intersection of $C_{nc} \cap C_{c}=r$. Using the identity
\begin{equation}
\frac{a\tau+b}{c\tau+d}=\frac{a}{c}-\frac{r}{c(c\tau+d)},
\end{equation}
with $ad-bc=r$, the total action becomes
\begin{equation}
S_{\text{bulk}}+S_{\text{bnd}}=-\pi i\frac{\tilde{k}_L}{2}\frac{a}{c}\, + \,\pi i \tilde{k}_L\frac{r}{c(c\tau+d)}\, - \,\pi i\frac{r\tilde{k}_L}{2}\frac{d}{c}\frac{1}{(c\tau+d)^2}.
\end{equation}
When $c=1$ and $a=d=0$ we must recover the on-shell action. In this case, the terms proportional to $a/c$ and $d/c$ vanish and we can compare with the on-shell entropy $S^{c=1}_{\text{on-shell}}$ defined in (\ref{on-shell entropy}). This means that more generally we have
\begin{equation}\label{on-shell action plus phases 1}
S_{\text{bulk}}+S_{\text{bnd}}=\frac{1}{c}S^{c=1}_{\text{on-shell}}-\pi i \frac{\tilde{k}_L}{2}\frac{a}{c}-\frac{\pi i}{2}r\tilde{k}_L\frac{d}{c}\frac{1}{(c\tau+d)^2},
\end{equation}
where the factor of $1/c$ in $S^{c=1}_{\text{on-shell}}$ is due to the orbifold. Therefore we identify $1/(c\tau+d)=-2i/(\phi^0)^{*}$ which remains constant for different $M_{(c,d)}$ geometries. Using the equation of motion for $\phi^0$, that is, $\hat{q}_0=\tilde{k}_L/(\phi^0)^2$, we compute the term proportional to $d/c$, 
\begin{equation}
-\frac{\pi i}{2}r\tilde{k}_L\frac{d}{c}\frac{1}{(c\tau+d)^2}=\frac{\pi i}{2}r\frac{d}{c}\frac{\Delta}{k_L},
\end{equation}
with $\hat{q}_0=\Delta/4k_L$, and $\Delta=Q^2P^2-(Q.P)^2$ and $k_L=P^2/2$; we have simplified the dependence on the matrix $D_{ab}$ for the $T^6$ and $K3\times T^2$ compactifications using the formulae (\ref{Cqq 1}) and (\ref{Cqq 2}). 
 
 In the presence of the singular fluxes  $f_a,\bar{f}_a$ with integer Chern-classes,  the on-shell entropy becomes, as explained in detail in \cite{Gomes:2017bpi},
\begin{equation}
S^{c=1}_{\text{on-shell}}=\pi r\hat{q}_0(\phi^0)^{*}+ \frac{\pi r}{(\phi^0)^{*}}\left(\tilde{k}_L(f)-2(\Delta f)^2\right),
\end{equation}
where 
\begin{equation}
6\tilde{k}_L(f)=p^3+p^a(c_{2a}-12(f_a+\bar{f}_a)),
\end{equation}
is the renormalized level, with $f_a,\bar{f}_a \in\mathbb{Z}^{+}$, which is directly related to the polarity. We have defined $\Delta f_a=f_a-\bar{f}_a$ and $(\Delta f)^2=\Delta f^a\Delta f^b D_{ab}$, with $\Delta f_a=D_{ab}\Delta f^b$. More explicitly, the polarity has the form
\begin{eqnarray}\label{polarity}
\frac{\tilde{k}_L(f)-2(\Delta f)^2}{4}&=&\frac{(P^2/2-(f_1-\bar{f}_1))^2}{2P^2}-\bar{f}_1+n_p\nonumber\\
&=&\frac{\nu^2}{4k_L}-m>0,
\end{eqnarray}
where we identified $\nu=P^2/2-(f_1-\bar{f}_1)$ and $m=\bar{f}_1-n_p$; here $n_p=0,1$ for the $T^4,K3$ CHL compactifications respectively. The final on-shell result, including the phases coming from the $SL(2,\mathbb{R})_L$ Chern-Simons terms, can be obtained from the expression (\ref{on-shell action plus phases 1}) substituting $\tilde{k}_L$ by $\tilde{k}_L(f)-2(\Delta f)^2$, that is, 
\begin{eqnarray}\label{on-shell action w phases}
S_{\text{bulk}}+S_{\text{bnd}}&=&\pi \frac{r}{c} \frac{\Delta}{4 k_L}(\phi^0)^{*}+4\pi\frac{r}{c} \frac{1}{(\phi^0)^{*}}\left(\frac{\nu^2}{4k_L}-m\right)-2\pi i \left(\frac{\nu^2}{4k_L}-m\right)\frac{a}{c}+\frac{\pi i}{2}r\frac{\Delta}{k_L}\frac{d}{c}\\
&=&2\pi \sqrt{\frac{\Delta}{k_L}\left(\frac{\nu^2}{4k_L}-m\right)}-2\pi i \left(\frac{\nu^2}{4k_L}-m\right)\frac{a}{c}+\frac{\pi i}{2}r\frac{\Delta}{k_L}\frac{d}{c}.
\end{eqnarray}

So far we studied the five dimensional theory and a  three dimensional effective Chern-Simons theory that one obtains after reduction on the sphere. In the full theory one expects additional $SU(2)_L$ Chern-Simons as explained in \cite{Dabholkar:2014ema}. However, at the level of the five dimensional theory we only see a $U(1)$ subgroup of this $SU(2)_L$. From the three dimensional point of view, the $SU(2)_L$ term has shown to give rise to the multiplier matrix that is part of the generalized Kloosterman sums. This was shown first for the Kloosterman sums of index one \cite{Dabholkar:2014ema}, which are relevant in the counting of $\CN=8$ dyons. More recently, this result has been extended for generalized Kloosterman sums of arbitrary index \cite{Gomes:2017bpi}, which appear in the counting of $\CN=4$ dyons. 

The analysis of the $SU(2)_L$ Chern-Simons that we folllow here is the same  as in \cite{Gomes:2017bpi} in almost every step, and so we refer the reader  to \cite{Gomes:2017bpi} for more details. An important difference, nevertheless,  regards the boundary terms of the Chern-Simons action, which as explained must be integrated $r$ times in this case. The Wilson lines of the flat connections around the cycles $C_1,C_2$ are exactly the same as in \cite{Gomes:2017bpi}. Hence the holonomies on the contractible and non-contractible cycles can be determined following the map (\ref{cycles map}).  The result of this computation is
\begin{equation}\label{SU2 CS phases}
I_{\text{CS+Bnd}}=\frac{\pi i}{2k_L}\frac{a}{c}\left(\nu+2n k_L\right)^2-\frac{\pi i}{k_L c}rQ.P(\nu+2k_L n)+\frac{\pi i}{2k_L}r\frac{d}{c}(Q.P)^2,
\end{equation}
where $I_{\text{CS+Bnd}}$ is the total Chern-Simons action including the boundary terms, and we have used the condition $ad-bc=r$ at an intermediate step. The integer $n$, which belongs to $\mathbb{Z}/c\mathbb{Z}$, parametrizes the Wilson line along the cycle $C_1$ and hence can be summed over due to the microcanonical boundary conditions we are using. The dependence on $Q.P$ comes from the component $A_2$ of the connection, which is kept fixed.

At the quantum level, the value of $\phi^0$ in the 
formula (\ref{on-shell action w phases}) fluctuates due to the localization mechanism described in \cite{Gomes:2017eac}. Therefore, to obtain the quantum result we can leave the value of $\phi^0$  off-shell in the expression (\ref{on-shell action w phases}). The full integral including the various phases is precisely the Bessel function dressed by the Kloosterman sums. That is, putting all the pieces together we find
\begin{eqnarray}\label{Kloos times Bessel partition function}
Z_{M_{(c,d)}}=&&\frac{1}{r^{b_2/2}c\sqrt{ck_L}}\sum_{\substack{0\leq a,d<c\\ad=r\text{ mod}(c)}}\exp{\left[-2\pi i \left(\frac{\nu^2}{4k_L}-m\right)\frac{a}{c}+\frac{\pi i}{2}r\frac{\Delta}{k_L}\frac{d}{c}\right]}\times \nonumber \\ 
&&\sum_{n=0}^{c-1}\exp{\left[ \frac{\pi i}{2k_L}\frac{a}{c}\left(\nu+2n k_L\right)^2-\frac{\pi i}{k_L c}rQ.P(\nu+2k_L n)+\frac{\pi i}{2k_L}r\frac{d}{c}(Q.P)^2\right]}\times \nonumber \\
&& \int_{\epsilon -i\infty}^{\epsilon +i\infty}\frac{dt}{t^{1+b_2/2}}\exp{\left[\frac{2\pi}{c\,t}r\frac{ \Delta}{k_L}+2\pi \frac{r}{c}\left(\frac{\nu^2}{4k_L}-m\right)t\right]}.
\end{eqnarray}
At an intermediate step we have integrated out the $\phi^a$'s, whose integrals are gaussian. The result of this integration is the factor of $ 1/r^{b_2/2}\sqrt{\text{det}(D_{ab})}\sim 1/r^{b_2}\sqrt{k_L}$. The factor $1/c\sqrt{c}$ in the first line, comes from a normalization of the measure as explained in \cite{Gomes:2017bpi}.  We are skipping details concerning the sign of the determinant of $D_{ab}$, which were, nevertheless, taken care in \cite{Gomes:2017eac}.

The formula (\ref{Kloos times Bessel partition function}) is still not in a form desirable to compare with the microscopic answer studied in  previous sections. Essentially, the term proportional to $ r (Q.P)^2$ in the second line of (\ref{Kloos times Bessel partition function}) does not have the canonical form $l^2 d/c$, with $l$ the coefficient that multiplies $\frac{\pi i}{k_L c}(\nu+2k_L n)$ in the same line, as suggested by the Kloosterman sums. It happens that we can rearrange the terms in (\ref{Kloos times Bessel partition function}) to obtain the desired answer. Note that $\Delta=4k_L n_0-(Q.P)^2$ with $n_0\in\mathbb{N}$, and so the term proportional to $(Q.P)^2 d/c$ in the second line cancels against a similar term coming from the the term $\Delta d/c$ in the first line. Therefore, we can add the term $ -\frac{\pi i}{2k_L}\frac{d}{c}(rQ.P)^2$ to the first line and subtract the same term on the second line. This gives
\begin{eqnarray}\label{Kloos times Bessel partition function 2}
Z_{M_{(c,d)}}=&&\frac{1}{r^{b_2/2}c\sqrt{ck_L}}\sum_{\substack{0\leq a,d<c\\ad=r\text{ mod}(c)}}\exp{\left[-2\pi i \left(\frac{\nu^2}{4k_L}-m\right)\frac{a}{c}+\frac{\pi i}{2}\frac{\Delta'}{k_L}\frac{d}{c}\right]} \nonumber \\
&&\times\sum_{n=0}^{c-1}\exp{\left[ \frac{\pi i}{2k_L}\frac{a}{c}\left(\nu+2n k_L\right)^2-\frac{\pi i}{k_L c}rQ.P(\nu+2k_L n)+\frac{\pi i}{2k_L}\frac{d}{c}(r Q.P)^2\right]} \nonumber\\ 
&& \times\int_{\epsilon -i\infty}^{\epsilon +i\infty}\frac{dt}{t^{1+b_2/2}}\exp{\left[\frac{2\pi}{c\,t}r\frac{ \Delta}{k_L}+2\pi\frac{r}{c}\left(\frac{\nu^2}{4k_L}-m\right)t\right]},
\end{eqnarray}
with $\Delta'=4(rn_0)k_L-(rQ.P)^2$. We can now identify the two first lines of the above expression with the modified version of the generalized Kloosterman sums. In particular, we have the following result
\begin{eqnarray}
\tilde{K}l(rn_0,rQ.P;m,\nu;k_L,c,r)&&=\frac{1}{\sqrt{ick_L}}\sum_{\substack{0\leq a,d<c\\ad=r\text{ mod}(c)}}\exp{\left[-2\pi i \left(\frac{\nu^2}{4k_L}-m\right)\frac{a}{c}+\frac{\pi i}{2}\frac{\Delta'}{k_L}\frac{d}{c}\right]} \nonumber\\ 
&&\times\sum_{n=0}^{c-1}\exp{\left[ \frac{\pi i}{2k_L}\frac{a}{c}\left(\nu+2n k_L\right)^2-\frac{\pi i}{k_L c}rQ.P(\nu+2k_L n)+\frac{\pi i}{2k_L}\frac{d}{c}(r Q.P)^2\right]} .\nonumber\\
{}
\end{eqnarray}

\subsubsection{$\CN=4$ Holography for Dyons with Torsion}

We apply the results of the previous section to compute the exact $AdS_2$ partition function for one-quarter BPS dyons with torsion, in four dimensional $\CN=4$ string theory compactifications. 

As shown in \cite{Dabholkar:2012nd}, the degeneracies of single center black holes in the $\CN=4$ theory correspond to the Fourier coefficients of a mock-modular Jacobi form. In this case, we can not apply the usual Rademacher expansion  to the Fourier coefficients. Due to the mock-modular properties of the partition function, the Rademacher expression contains additional corrections  \cite{Ferrari:2017msn}. These corrections, however, occur when the polarity attains its minimum value, and so for very large charges one can show that they contribute at much subleading order. Though our analysis does not take these corrections into account, for the purpose of demonstrating the arithmetic properties of the degeneracy it is enough to consider the generalized Kloosterman sums as if we were dealing with an honest Jacobi form. As a matter of fact, the Kloosterman sums, in particular, the dependence on the multiplier matrix, have their origin in the modular transformations of the theta functions, which are used to write the mock Jacobi forms in a Jacobi-theta expansion. It follows from this that all the terms in Rademacher expansion of the mock-modular form are dressed by the usual generalized Kloosterman sums \cite{Ferrari:2017msn}. Therefore, if our analysis included also the corrections coming from the mock nature of the partition function, one would expect similar arithmetic properties.

The full answer for the partition function is a sum over various terms. For each configuration parameterized by $(f,\bar{f})\sim (\nu,m)$, each term in the partition function is the product of $\Omega(\nu,m)$, which depends only on the Calabi-Yau data and parametrizes fluctuations of the internal geometry with the fluxes $f,\bar{f}$ turned on, and a term corresponding to the supergravity part that we denote by $Z_{AdS_2\times S^1\times S^2/\mathbb{Z}_c}(\nu,m)$. Summing over arbitrary $\mathbb{Z}_c$ orbifolds we have 
\begin{equation}\label{Z_r for N=4}
Z_r=\sum_{\nu,m}\Omega(\nu,m)\sum_{c=1}^{\infty}Z_{AdS_2\times S^1\times S^2/\mathbb{Z}_c}(\nu,m).
\end{equation}
$Z_{r}$ is the partition function for the geometries that obey $ad-bc=r$. For $r=1$, we can compute $\Omega(\nu,m)$  from gravity \cite{Gomes:2017eac} and match the microscopic analysis derived from the Siegel modular forms \cite{Gomes:2015xcf,Murthy:2015zzy}. Since $Z_{AdS_2\times S^1\times S^2/\mathbb{Z}_c}(\nu,m)$ contains the dependence on the Kloosterman sums and the Bessel function, we can identify  $\Omega(\nu,m)$ with the polar coefficients of the microscopic answer \cite{Gomes:2017eac}.  It is important to stress once more that in the expression (\ref{Z_r for N=4}) all the information that enters in the computation of $\Omega(\nu,m)$ is determined by the theory with $r=1$.

 We have found in the previous section that the expression for the partition function $Z_{AdS_2\times S^1\times S^2/\mathbb{Z}_c}(\nu,m)$ is the product of the Bessel function times the modified generalized Kloosterman sums. The full partition function $Z_r$  becomes
\begin{eqnarray}\label{N=4 partition fnct 1}
Z_r=&&\frac{1}{r^{b_2/2}}\sum_{\nu,m}\Omega(\nu,m)\sum_{c=1}^{\infty}\frac{1}{c} \tilde{K}l(rn, rQ.P;m,\nu;k_L,c,r) \nonumber\\
&&\times\int_{\epsilon-i\infty}^{\epsilon+i\infty}\frac{dt}{t^{1+b_2/2}}\exp{\left[\frac{2\pi}{t}\frac{r}{c}\frac{ \Delta}{k_L}+2\pi \frac{r}{c}\left(\frac{\nu^2}{4k_L}-m\right)t\right]},
\end{eqnarray}
with $\Delta=4k_Ln-(Q.P)^2$. In this expression, the variables $\nu,m,k_L$ are all determined by the theory with $r=1$. 

Now we can use the Selberg identity for the modified generalized Kloosterman sum (\ref{modified Selberg id gen Kloos 1}),
\begin{equation}
\tilde{Kl}(rn,rQ.P;m,\nu;k_L,c,r)=\sum_{s|(c,r)}s^{3/2} Kl(nr^2/s^2,rQ.P/s;m,\nu;k_L,c/s),
\end{equation}
where we used that $\text{gcd}(c,nr,rQ.P,r)=\text{gcd}(c,r)$. So we can write 
 \begin{eqnarray}
 Z_r=&&\frac{1}{r^{b_2/2}}\sum_{\nu,m}\Omega(\nu,m)\sum_{c=1}^{\infty}\frac{1}{c}\sum_{s|(c,r)}s^{3/2} Kl(nr^2/s^2,rQ.P/s;m,\nu;k_L,c/s)\nonumber\\  &&\times\int_{\epsilon-i\infty}^{\epsilon+i\infty}\frac{dt}{t^{1+b_2/2}}\exp{\left[\frac{2\pi}{c\,t}r\frac{ \Delta}{k_L}+2\pi \frac{r}{c}\left(\frac{\nu^2}{4k_L}-m\right)t\right]}.
 \end{eqnarray}
 Then, for each $s$ term in the sum above, we rescale  $t\rightarrow ts/r$ in the integral to obtain
\begin{eqnarray}
Z_r=&&\sum_{\nu,m}\Omega(\nu,m)\sum_{c=1}^{\infty}\sum_{s|(c,r)}\frac{s^{1/2-b_2/2}}{c/s} Kl(nr^2/s^2,rQ.P/s;m,\nu;k_L,c/s) \nonumber\\  &&\times\int_{\epsilon-i\infty}^{\epsilon+i\infty}\frac{dt}{t^{1+b_2/2}}\exp{\left[\frac{2\pi}{c/s\,t}r^2/s^2\frac{ \Delta}{k_L}+\frac{2\pi}{c/s}\left(\frac{\nu^2}{4k_L}-m\right)t\right]}.
\end{eqnarray}
Finally, interchanging the sums over $c$ and $s$, using the property (\ref{sum c sum s}), we find
\begin{eqnarray}
Z_r&=&\sum_{s|r}s^{1/2-b_2/2}\sum_{\nu,m}\Omega(\nu,m)\sum_{c=1}^{\infty}\frac{1}{c} Kl(nr^2/s^2,rQ.P/s;m,\nu;k_L,c) \nonumber\\  &&\times\int_{\epsilon-i\infty}^{\epsilon+i\infty}\frac{dt}{t^{1+b_2/2}}\exp{\left[\frac{2\pi}{c\,t}r^2/s^2\frac{ \Delta}{k_L}+\frac{2\pi}{c}\left(\frac{\nu^2}{4k_L}-m\right)t\right]}\nonumber\\
{}\nonumber\\
&=&\sum_{s|r}s^{1/2-b_2/2}d_{r=1}(Q'=rQ/s,P) .\label{Hecke N=4}
\end{eqnarray}
For each $s$, we sum over  $c\geq 1$  and $\nu,m$. This gives precisely the Rademacher expansion of the primitive degeneracy $d_{r=1}$, modulo the mock-modular corrections. Furthermore, we have $1/2-b_2/2=\omega-1$, where $\omega$ is the weight of the mock-modular form, and so we can identify  the expression (\ref{Hecke N=4}) with the action of the Hecke operator $V_r$ on the primitive degeneracy, with the electric charges rescaled by a factor of $r$. 

We have assumed, from the beginning, that both $Q$ and $P$ were primitive vectors with $\text{gcd}(Q\wedge P)=1$. From the formula above we have
\begin{equation}
Q'=r Q,
\end{equation}
 and so it becomes clear that the electric charges of the near-horizon geometry are effectively rescaled by a factor of $r$.
 This means that the torsion invariant as measured from the near-horizon \footnote{By measured at the near-horizon, we mean the particular quantities that appear in the $AdS_2$ renormalized action.} is
 \begin{equation}
\text{gcd}(Q'\wedge P)=r.
 \end{equation}
 Hence, we find that, apart from the power of $s$, the expression (\ref{Hecke N=4}) has precisely the structure predicted by the non-primitive formula (\ref{non-primitive N=4}).

\subsubsection{$\CN=8$ Holography for Non-primitive Dyons}

In this section, we consider the computation of the $AdS_2$ partition function of one-eighth BPS dyons with non-primitive charges in $\CN=8$ string theory compactifications.  We follow similar steps as in the $\CN=4$ example studied previously. However, in this case we will impose different quantization conditions on the electric charges at the horizon. Imposing different quantization conditions will allow us to generate dependence on additional arithmetic invariants. We will show that this dependence is in perfect agreement with the microscopic answers discussed in section \S \ref{sec non-primitive N=8}. 

As before, we will be summing over geometries $M_{(c,d)}$ with $ad-bc=r$, but in this case we impose that the electric charges at the horizon are multiples of $r_2$ with 
\begin{equation}
r=r_1r_2,\;\; \text{gcd}(r_1,r_2)=1 .
\end{equation}
We will comment on the case for which $r_1,r_2$ have common factors. As explained previously, due to the orbifold geometry, the chemical potentials $\phi^{0,a}$ appear multiplied by a factor of $r=r_1r_2$ at the horizon. In the $\CN=8$ case, we impose that the charges are quantized as $q_{0,a}\rightarrow q_{0,a}/r_1$, which implies that at the horizon the electric charges are proportional to $r_2$. That is, we have
\begin{equation}
q_{0,a}'=r_2q_{0,a},\;\;q_{0,a}\in \mathbb{Z},
\end{equation}
where $q_{0,a}'$ are the charges that couple to the potentials $\phi^{0,a}$. Furthermore, we must enforce that
\begin{equation}
 P^2/2=k_L=1,
\end{equation}
 which is the level observed in the microscopic answer. Although we impose $k_L=1$, this analysis is different from the approach followed in \cite{Dabholkar:2011ec,Dabholkar:2014ema}. In particular, the polarity is proportional to $r$ and it is thus not restricted to finite values as in \cite{Dabholkar:2014ema}. This means that we can have arbitrarily small values of $1/\phi^0\sim 1/\sqrt{rk_L}$ by taking $r\gg 1$, which is in contrast with the situation in \cite{Dabholkar:2014ema} where $1/\phi^0$ is of order $1/\sqrt{k_L}$. Having small values of $1/\phi^0$ means that the M-theory radius, for unit size $AdS_2$, can be arbitrarily small and so we can have a four dimensional weakly coupled description of the black hole as explained in \cite{Gomes:2017eac}. 
 
 Since the electric charges are rescaled by $r_2$ instead, we have
 \begin{equation}\label{N=8 charge rescaling}
 r(q_0-D^{ab}q_aq_b/2)\rightarrow r_2 \left(q_0-\frac{D^{ab}}{2r_1}q_aq_b\right),
 \end{equation}
 after $q_{0,a}\rightarrow q_{0,a}/r_1$. Furthermore we enforce that
 \begin{equation}
 -q_0+\frac{C^{ij}}{2r_1}q_iq_j\equiv n,\;n\in \mathbb{Z},
 \end{equation}
 that is, the condition that $Q^2/2=n$ (\ref{Cqq 1}). So we can write
 \begin{equation}
  -r(q_0-D^{ab}q_aq_b/2)\rightarrow r_2\left(n-\frac{(Q.P)^2}{4k_Lr_1}\right)=r_2n-\frac{(r_2Q.P)^2}{4k_L r}.
 \end{equation}
 With these charges, we substitute $n\rightarrow r_2n$ and $l\rightarrow r_2Q.P$ in the modified Kloosterman sum $\tilde{K}l(n, l;m,\nu;k_L,c,r)$. Using its arithmetic property, we find 
\begin{eqnarray}\label{mod Kloos N=8 bulk}
\tilde{K}l(r_2n, r_2Q.P;m,\nu;k_L,c,r)&=&\sum_{s|(c,r,r_2n,r_2Q.P)}s^{3/2} Kl(r_2nr/s^2,r_2Q.P/s;m,\nu;k_L,c/s)\nonumber\\
&=&\sum_{s|(c,r_1r_2,r_2n,r_2Q.P)}s^{3/2} Kl(r_2nr/s^2,r_2Q.P/s;m,\nu;k_L,c/s).\nonumber\\
{}
\end{eqnarray}

The full partition is again a sum over the Bessel functions multiplied by the modified Kloosterman sums. In the $\CN=8$ problem only the most polar term with $\nu=\nu_0=k_L=1$ and $m=m_0=0$ contributes as shown in \cite{Gomes:2017eac}. We plug the formula (\ref{mod Kloos N=8 bulk}) in the full partition function (\ref{Z_r for N=4}). At an intermediate step we rescale $t\rightarrow ts/r$ such that the degeneracy becomes
\begin{eqnarray}\label{N=8 non-deg bulk 1}
Z_r=&&\Omega(\nu_0,m_0)\sum_{c=1}^{\infty}\sum_{s|(c,r_1r_2,r_2n,r_2Q.P)}\frac{s^{1/2-b_2/2}}{c/s} Kl(r_2nr/s^2,r_2Q.P/s;m_0,\nu_0;k_L,c/s) \nonumber\\  
&&\times\int_{\epsilon-i\infty}^{\epsilon+i\infty}\frac{dt}{t^{1+b_2/2}}\exp{\left[\frac{2\pi}{c/s\,t}\frac{ r_2^2r_1\Delta'/s^2}{k_L}+\frac{2\pi}{c/s}\left(\frac{\nu_0^2}{4k_L}-m_0\right)t\right]},
\end{eqnarray}
with 
\begin{equation}\label{Delta'}
\frac{\Delta'}{k_L}=4n-\frac{(Q.P)^2}{r_1k_L}.
\end{equation}
 This gives in addition, 
 \begin{equation}
 r_2^2r_1\Delta'=4(r_2^2n)(r_1k_L)-(r_2Q.P)^2.
 \end{equation}

Note that for large charges the entropy must go as
 \begin{equation}
 \sim \pi \sqrt{Q'^2P'^2-(Q'.P')^2},\;Q',P'\gg 1,
 \end{equation}
 for some charges $Q'$ and $P'$; this is a consequence of duality invariance. From the expression (\ref{N=8 non-deg bulk 1}), the Bessel grows exponentially with
 \begin{equation}
 \sim \pi \sqrt{r_2^2r_1 \Delta'},
 \end{equation}
 and so we must have 
 \begin{equation}
 r_2^2r_1\Delta'=Q'^2P'^2-(Q'.P')^2.
 \end{equation}
Therefore, at the horizon the effective charges $(Q',P')$ obey 
 \begin{equation}
 Q'^2/2=r_2^2n,\;P'^2/2=r_1k_L,\; Q'.P'=r_2Q.P .
 \end{equation}
 From the expression (\ref{N=8 charge rescaling}), we see that by setting $D^{ab}/r_1\equiv D'^{ab}$, and thus $D'_{ab}=r_1D_{ab}$, we have $P'^2=C'_{ij}p^ip^j=r_1k_L$, with $C'_{ij}=r_1C_{ij}$.
 Moreover, since $Q'^2$ and $Q'.P'$ are proportional, respectively, to $r_2^2$ and $r_2$ we have $Q'=r_2Q'^{0}$, with $Q'^{0}$ a primitive vector. Similarly, $P'$ must be primitive because neither $P'^2$ nor $Q'.P'$ are proportional to $I^2$ and $I$ respectively, with $I=\text{gcd}(P')$. To be more explicit, let us consider the following charge vectors in a four dimensional subspace of the eight dimensional lattice of $SO(4,4)\subset SO(6,6)$ of the T-duality group,
 \begin{equation}\label{N=8 dyon charges}
 Q'=r_2\left(\begin{array}{c}
 1\\
 n\\
 0\\
 0
 \end{array}\right),\;P'=\left(\begin{array}{c}
 0\\
 Q.P\\
 1\\
 r_1k_L
 \end{array}\right).
 \end{equation}
 We can take the metric in this subspace to be
 \begin{equation}
 L=\left(\begin{array}{cccc}
 0 & 1 & 0 & 0\\
 1 & 0 & 0 &0\\
 0 & 0 & 0 &1 \\
 0 & 0 & 1 &0
 \end{array}\right).
 \end{equation}
 The T-duality invariant combinations are therefore 
 \begin{equation}
 Q'^2=Q'^T L Q'= 2 r_2^2 n,\;\;P'^2=P'^TLP'=2r_1k_L,\;\;Q'.P'=Q'^TLP'=r_2Q.P.
 \end{equation}
 It is also easy to see that we have 
 \begin{equation}
 \text{gcd}(Q'\wedge P')=r_2.
 \end{equation}
 
 Interchanging the sums over $c$ and $s$ in (\ref{N=8 non-deg bulk 1}) we finally obtain
\begin{eqnarray}\label{N=8 non-prim bulk}
Z_r&=&\Omega(\nu_0,m_0)\sum_{s|r_2\text{gcd}(Q'^2/2,P'^2/2,Q'.P')}s^{1/2-b_2/2}\sum_{c=1}^{\infty}\frac{1}{c} Kl(Q'^2P'^2/4s^2,Q'.P'/s;m_0,\nu_0;k_L,c)\times \nonumber\\  &&\int_{\epsilon-i\infty}^{\epsilon+i\infty}\frac{dt}{t^{1+b_2/2}}\exp{\left[\frac{2\pi}{c\,t}\frac{ \left(Q'^2P'^2-(Q'.P')^2\right)}{s^2k_L}+\frac{2\pi}{c}\left(\frac{\nu_0^2}{4k_L}-m_0\right)t\right]}\nonumber\\
&&{}\nonumber\\
&=&\sum_{s|\text{gcd}(Q'\wedge P')\text{gcd}(Q'^2/2,P'^2/2,Q'.P')}s^{1/2-b_2/2}\,d_{r=1}(Q'^2P'^2/4s^2,Q'.P'/s),
\end{eqnarray}
where we have used the fact that 
\begin{equation}
\text{gcd}(n,Q.P,r_1)=\text{gcd}(Q'^2/2,P'^2/2,Q'.P'),
\end{equation}
provided that $\text{gcd}(r_1,r_2)=1$. Formula (\ref{N=8 non-prim bulk}) captures the degeneracy for dyons of the form (\ref{N=8 dyon charges}). Since we have $\text{gcd}(Q'\wedge P')=r_2$, the condition $\text{gcd}(r_1,r_2)=1$  is also equivalent to 
\begin{equation}
\text{gcd}(P'^2/2,\text{gcd}(Q'\wedge P'))=1\Leftrightarrow\text{gcd}\Big(Q'^2/2,P'^2/2,Q'.P',\text{gcd}(Q'\wedge P')\Big)=1.
\end{equation}
This condition is precisely the primitivity condition imposed by $\psi(q)=1$ in (\ref{primitivity cond 1}). With the exception of  the power of $s$ that multiplies the primitive degeneracy $d_{r=1}$, we obtain precise agreement with the microscopic answer (\ref{duality deg}) including the primitivity conditions (\ref{primitivity cond 1}).

We can try to generalize the discussion for the case with $(r_1,r_2)$ not co-prime. This means that we are relaxing the condition $\psi(q)=1$, which is one of the $E_{7,7}(\mathbb{Z})$ discrete invariants. In this case, we propose that  to obtain a duality invariant formula, we need to supplement the answer (\ref{N=8 non-deg bulk 1}) with additional geometries. At the moment we do not have a physical understanding of these new contributions, but simply show that they lead to the desired result. Our proposal consists in introducing an additional sum over the divisors of $\text{gcd}(r_1,r_2)$. In particular, we propose
\begin{eqnarray}
Z_{(r_1,r_2)}=&&\sum_{d|(r_1,r_2)}d^{1/2-b_2/2}Z_{r/d^2}\nonumber\\
=&&\Omega(\nu_0,m_0)\sum_{c=1}^{\infty}\sum_{d|(c,r_1,r_2)}d^{1/2-b_2/2}Z_{AdS_2\times S^1\times S^2/\mathbb{Z}_{c/d}}(r/d^2),
\end{eqnarray}
with $Z_{r/d^2}$ the expression for $(r_1,r_2)$ co-prime. In this case, it looks like we are summing over configurations with different number of $\D6-\aD6$ branes, which is parameterized by the integer $r/d^2$. As in the co-prime case, we have only one polar term. The factor $d^{1/2-b_2/2}$ is introduced to obtain the necessary duality invariant result. More explicitly we have
\begin{eqnarray}
Z_{(r_1,r_2)}=&&\Omega(\nu_0,m_0)\sum_{c=1}^{\infty}\sum_{d|(c,r_1,r_2)}\frac{d^{1/2-b_2/2}}{c/d}\tilde{K}l(r_2n, r_2Q.P;m_0,\nu_0;k_L,c/d,r/d^2)\nonumber\\
&&\times d^{b_2}\int_{\epsilon-i\infty}^{\epsilon+i\infty}\frac{dt}{t^{1+b_2/2}}\exp{\left[\frac{2\pi}{c/d\,t}r_2\frac{ \Delta'}{k_L}+\frac{2\pi}{c/d}\frac{r}{d^2}\left(\frac{\nu_0^2}{4k_L}-m_0\right)t\right]},
\end{eqnarray}
with  $\Delta'=4k_Ln-(Q.P)^2/r_1$. The factor of $d^{b_2}$ in the second line is due to the gaussian integration over the variables $\phi^{a}$, with $a=1\ldots b_2$. We have rescaled the electric charges $q_{0,a}$ in such way that $\hat{q}_0r/d^2$ is proportional to $r_2\Delta'/k_L$ and becomes  independent of $d$ throughout the computation. We have defined $n\equiv-q_0+C^{ij}q_iq_j/2r_1$ and $Q.P\equiv q_ip^i$.  This means that we have rescaled the charges $q$ as 
\begin{equation}\label{rescaling d}
q_{0}\rightarrow q_{0}d^2/r_1,\;q_a\rightarrow q_a d/r^1.
\end{equation}
Note the in-homogeneous re-scaling of the charges.  The rescaling (\ref{rescaling d})  is a bit unnatural because it means that we are changing the charges for each term in the sum over $d$. We do not understand this fact but it will lead to the desired answer.

We can show that the modified Kloosterman sums and Bessel function follow from the reasoning discussed in section \S \ref{sec mod Kloos CS}, but now for the orbifold with $c\rightarrow c/d$ and $r\rightarrow r/d^2$.  We develop the modified Kloosterman as a sum of Kloosterman sums using formula (\ref{mod Kloos N=8 bulk}), and hence, following the steps explained before, we obtain
\begin{eqnarray}
Z_{(r_1,r_2)}=&&\Omega(\nu_0,m_0)\times\nonumber\\
&&\times\sum_{c=1}^{\infty}\sum_{d|(c,r_1,r_2)}\sum_{s|\left(\frac{c}{d},\frac{r_1r_2}{d^2},r_2n,\frac{r_2Q.P}{d}\right)}\frac{d^{1/2+b_2/2}s^{3/2}}{c/d} Kl(r_2nr/(ds)^2,r_2Q.P/ds;m_0,\nu_0;k_L,c/ds) \nonumber\\  
&&\times\int_{\epsilon-i\infty}^{\epsilon+i\infty}\frac{dt}{t^{1+b_2/2}}\exp{\left[\frac{2\pi}{c/d\,t}r_2\frac{ \Delta'}{k_L}+\frac{2\pi}{c/d}\frac{r}{d^2}\left(\frac{\nu_0^2}{4k_L}-m_0\right)t\right]}.
\end{eqnarray}
After rescaling $t\rightarrow t d^2s/r$ we get
\begin{eqnarray}
Z_{(r_1,r_2)}=&&\Omega(\nu_0,m_0)\times\nonumber\\
&&\times\sum_{c=1}^{\infty}\sum_{d|(c,r_1,r_2)}\sum_{s|\left(\frac{c}{d},\frac{r_1r_2}{d^2},r_2n,\frac{r_2Q.P}{d}\right)}\frac{(ds)^{1/2-b_2/2}}{c/sd} Kl(r_2nr/(ds)^2,r_2Q.P/ds;m_0,\nu_0;k_L,c/ds) \nonumber\\  
&&\times\int_{\epsilon-i\infty}^{\epsilon+i\infty}\frac{dt}{t^{1+b_2/2}}\exp{\left[\frac{2\pi}{c/ds\,t}\frac{r_2^2r_1}{(sd)^2}\frac{ \Delta'}{k_L}+\frac{2\pi}{c/ds}\left(\frac{\nu_0^2}{4k_L}-m_0\right)t\right]}.
\end{eqnarray}
We write further,
\begin{eqnarray}
Z_{(r_1,r_2)}=&&\Omega(\nu_0,m_0)\sum_{\substack{a=sd\\d|(r_1,r_2)\\s|(r_1r_2/d^2,r_2n,r_2Q.P/d)}}a^{1/2-b_2/2}N(a)\sum_{c=1}^{\infty}\frac{1}{c} Kl(r_2^2r_1n/a^2,r_2Q.P/a;m_0,\nu_0;k_L,c) \nonumber\\  
&&\times\int_{\epsilon-i\infty}^{\epsilon+i\infty}\frac{dt}{t^{1+b_2/2}}\exp{\left[\frac{2\pi}{c\,t}\frac{r_2^2r_1}{a^2}\frac{ \Delta'}{k_L}+\frac{2\pi}{c}\left(\frac{\nu_0^2}{4k_L}-m_0\right)t\right]}\nonumber\\
{}\nonumber\\
&&=\sum_{\substack{a=sd\\d|(r_1,r_2)\\s|(r_1r_2/d^2,r_2n,r_2Q.P/d)}}a^{1/2-b_2/2}N(a)\,d_{r=1}(r_2^2r_1n/a^2,r_2Q.P/a),
\end{eqnarray}
where we have interchanged the sums over $c$ and $a$. The function $N(a)$, which was defined in (\ref{number divisors}), counts the number of divisors $\delta$, with $\delta$ defined by $d=a\delta/(a,r_2)$,
\begin{equation}
N(a)=\text{number of divisors }\delta\,\left(r_1,r_2,r_2n,a,\frac{r_1r_2}{a},\frac{r_1r_2n}{a},\frac{r_2^2n}{a},\frac{r_2^2r_1n}{a^2}\right),
\end{equation}
with the condition that $N(a)=0$ unless $a|(r_1r_2,r_2^2n,r_1r_2n)$ and $a^2|r_2^2r_1n$.

\subsection{Dabholkar-Harvey states and U-duality}

In this section, we explore the degeneracy of one-half BPS states in $\CN=4$ string theory and its dependence on the arithmetic invariants. To our knowledge, a microscopic understanding of the counting is not known in the non-primitive case. Nevertheless,  using our bulk methods we will be able to provide such a formula.

When the electric and magnetic charge vectors of a $\CN=4$ dyon are proportional to each other, the state preserves additional supersymmetries and gives rise to a half-BPS dyon. After an electric-magnetic duality transformation, we can bring the dyon to a purely electric form, which we can interpret as perturbative momentum-winding states in the Heterotic frame- these are the well known Dabholkar-Harvey states \cite{Dabholkar:1989jt,Dabholkar:2005dt}. The counting of such states is captured by the Dedekind-Ramanujan function as follows
\begin{equation}\label{1/2 BPS dyons}
\sum_{m}d_{\text{D-H}}(m)q^m=\frac{1}{\eta^{24}(q)}=\frac{1}{q\prod_{n=1}^{\infty}(1-q^n)^{24}},
\end{equation}
where $d_{\text{D-H}}(m)$ is the degeneracy of a state with momentum $n$ and winding $w$ with $m=nw$. 

Lets consider a dyon with charges
\begin{equation}
(Q,P)=(rQ',sQ'),\;\text{gcd}(r,s)=1.
\end{equation}
Then, there is a $SL(2,\mathbb{Z})$ electric-magnetic duality transformation that brings the dyon to a purely electric configuration of the type $(Q',0)$. In this case, the only discrete invariant associated with this state is the multiplicity defined by
\begin{equation}
M=\text{gcd}(rQ',sQ')=\text{gcd}(Q').
\end{equation}

On general grounds, we expect the degeneracy of a such dyon to depend on the multiplicity $M$. In this section we provide such a formula derived purely from the gravitational theory. Our focus is on the Kloosterman sums, which as we have learned in the previous sections, contain information about the arithmetic invariants. At this stage we do not to fully understand how to reproduce the exact microscopic answer (\ref{1/2 BPS dyons}) from the bulk theory. From the analysis of \cite{Gomes:2015xcf,Dabholkar:2014ema,Gomes:2017eac} we can conclude, nevertheless, that the degeneracies $d_{\text{D-H}}$ have similar expansions as infinite sums of Bessel functions dressed by classical Kloosterman sums. In particular, from the analysis of \cite{Gomes:2017eac} we can predict that there is only one polar term in this expansion, which is in agreement with the microscopic counting (\ref{1/2 BPS dyons}).  However, the analysis of \cite{Gomes:2015xcf} fails to predict the exact index of the Bessel function that appears in the Rademacher expansion. 

The map of a purely electric state in the Heterotic frame to the M-theory frame implies that we have $p^a=0$ for $a\neq 1$. The precise map consists of the following: the winding $\omega$ is mapped to $p^1=\omega$ M5-branes wrapping $K3\times S^1$ and the momentum of the Heterotic string is mapped to momentum along the circle $S^1$. For our purposes we can consider a two charge configuration with electric charge $q_0$ and magnetic charge $p^1$. In the $\D6$ picture we have a configuration with $r \D6-r\aD6$ with total internal flux $p^1=r$, which means that after uplift we have $p^{1}/r=1$ units of G-flux. From the polarity condition (\ref{polarity}) we must conclude that $f_1=\bar{f}_1=0$ since we have $P^2=0$. This is also the condition imposed by the positivity of the metric \cite{Gomes:2017eac}. For such fluxes the metric becomes string scale size.

 Since the fluxes $f,\bar{f}$ must be turned off, we do not have to sum over spectral flow sectors. Moreover it implies that we have only one polar term and its coefficient is a order one term which we can normalize to one. In this case, the Kloosterman sums arise purely from the $SL(2,\mathbb{R})_L$ sector of the effective Chern-Simons theory \cite{Dabholkar:2014ema} and so we find classical rather than generalized Kloosterman sums. On the $\mathbb{Z}_c$ orbifold with $ad-bc=r$, the sum over flat connections gives rise to
\begin{equation}
\tilde{Kl}(rq_0,n_p,r,c)=\sum_{\substack{0\leq a,d<c \\ ad=r\text{ mod}(c)}}\exp{\left(-2\pi i n_p\frac{a}{c}+2\pi i r q_0, \frac{d}{c}\right)}
\end{equation}
which is the modified classical Kloosterman sum defined in (\ref{modified Kloos}). Here $n_p=1$ is the polarity of $1/\eta(q)^{24}=q^{-n_p}+\mathcal{O}(q^{0})$. The modified classical Kloosterman sums obey 
\begin{eqnarray}\label{sum Kloos 1/2BPS}
\tilde{Kl}(rq_0,n_p,r,c)&=&\sum_{s|(c,r,rq_0)}s\,Kl(r^2q_0/s^2,n_p,c/s)\nonumber\\
&=&\sum_{s|(c,r)}s\,Kl(r^2q_0/s^2,n_p,c/s).
\end{eqnarray}
We will use this property to produce a formula for the degeneracy dependent on the multiplicity.

To determine the Bessel function we follow the same steps as implemented in the previous examples. The quantum entropy gets multiplied by a factor of $r/c$. Integrating over $\phi^a$ in the quantum entropy with the measure of \cite{Gomes:2015xcf} leads to factors of $t$ in the Bessel integral. However, such measure does not reproduce the index $\nu=13$ expected for the Bessel function. We will not attempt to solve this puzzle here, and we leave the index $\nu$ undetermined. The full answer is 
\begin{equation}\label{DH rademacher 1}
d(q_0)_r=\sum_{c=1}^{\infty}\frac{1}{c}\tilde{Kl}(rq_0,n_p,r,c)\int_{\epsilon-i\infty}^{\epsilon +i\infty}\frac{dt}{t^{1+\nu}}\exp{\left(2\pi \frac{rq_0}{c\,t}+2\pi n_p\frac{r}{c}t\right)}.
\end{equation}
The factor $1/c$ multiplying the Kloosterman sum is due to the fact that we are summing over the flat connections parameterized by $a,d\in \mathbb{Z}/c\mathbb{Z}$; the factor of $1/c$ ensures that the measure in the path integral is well normalized \cite{Gomes:2017bpi}. Moreover, we have omitted an overall factor depending on $r$ that multiplies the expression (\ref{DH rademacher 1}); this comes from performing the gaussian integrals over $\phi^a$. We now use the Selberg identity for the modified Kloosterman sum (\ref{sum Kloos 1/2BPS}) and rescale $t\rightarrow t s/r$ to obtain
\begin{equation}
d(q_0)_r=r^{\nu}\sum_{c=1}^{\infty}\sum_{s|(c,r)}\frac{s^{-\nu}}{c/s}\,K(r^2q_0/s^2,n_p,c/s)\int_{\epsilon-i\infty}^{\epsilon +i\infty}\frac{dt}{t^{1+\nu}}\exp{\left(2\pi \frac{r^2q_0/s^2}{c/s\,t}+2\pi n_p \frac{t}{c/s}\right)}.
\end{equation}
We interchange the sums over $c$ and $s$ to find
\begin{eqnarray}
d(q_0)_r&=&r^{\nu}\sum_{s|r}s^{-\nu}\sum_{c=1}^{\infty}\frac{1}{c}\,K(r^2q_0/s^2,n_p,c)\int_{\epsilon-i\infty}^{\epsilon +i\infty}\frac{dt}{t^{1+\nu}}\exp{\left(2\pi \frac{r^2q_0/s^2}{c\,t}+2\pi n_p\frac{t}{c}\right)}\\
&=&r^{\nu}\sum_{s|r}s^{-\nu}d_{r=1}(r^2q_0/s^2).
\end{eqnarray}
The primitve answer $d_{r=1}$ is a function of $Q^2/2$, with $Q$ the electric charge vector measured at the horizon, and so  we can identify $Q^2/2=r^2q_0=n\omega$. Therefore, we find that $r$ is the multiplicity $M$ of the charge configuration.

\section{Non-Holomorphic Hecke Operators}\label{sec Hecke}

In the previous sections, we considered the bulk computation of the black hole degeneracy for charges with non-primitive factors. We have found that by including in the path integral geometries 
$M_{(c,d)}$, with $(c,d)$ not necessarily co-prime, we could reproduce the structure of the non-primitive answers as a sum over the primitive degeneracy formulas. Essentially, this happened as a consequence of the non-trivial properties of the generalized Kloosterman sums for non-primitive charges. Nevertheless, the agreement between the bulk and the microscopic answers was only up to a power of $s$, the integer divisor of the U-duality invariant. As a matter of fact, we were able to identify the full partition function with the action of an Hecke operator on the primitive answer. In this section, we try to address this puzzle and lay down some ideas that may solve the discrepancy observed. 

The main idea of this section is to write the $\CN=8$ and $\CN=4$ non-primitive microscopic degeneracies in a way that they can be related to Hecke operators acting on the primitive answers. Being able to write the microscopic degeneracy in this form can also be very useful, if we want to generalize those formulas to other U-duality invariants. The idea is to explore the action of multiple Hecke operators, which usually gives rise to more intricate dependence on the charges as exemplified in \cite{Maldacena:1999bp}. In particular, we review the discussion of section 6.1 in that paper. 

Let us first focus on the $\CN=8$ case; the $\CN=4$ case will follow by a simple generalization. To simplify the discussion, we consider charge configurations with $\text{gcd}(Q\wedge P)=1$. In this case, the degeneracy is given by
\begin{equation}\label{deg N=8 sec Hecke}
d(Q,P)=(-1)^{Q.P}\sum_{s|\text{gcd}(Q^2/2,P^2/2,Q.P)}s\,c(Q^2P^2/4s^2,Q.P/s),
\end{equation}
with $c(n,l)$ the Fourier coefficient of $\phi_{-2,1}(\tau,z)$, which is a Jacobi form with weight $\omega=-2$ and index $1$. The structure of this formula is very similar to the action of the  Hecke operator $V_{P^2/2}$ acting on the Fourier coefficients of $\phi_{-2,1}(\tau,z)$ \cite{Eichler:1985ja}. However, the power of $s$ has a non-standard value, which should be $s^{\omega-1}=s^{-3}$, because $\phi_{-2,1}(\tau,z)$ has weight $-2$. 

The fact that the formula (\ref{deg N=8 sec Hecke}) does not have the Hecke operator form is problematic. If we reverse the steps that take us from the Hecke operator acting on the modular form to the expression for the Fourier coefficients, we find that the final object does not transform correctly under modular transformations. Despite this, one can show that (\ref{deg N=8 sec Hecke}) is the Fourier coefficient of the modified elliptic genus of a partition function that is a Hecke operator acting on the primitive answer, which is modular. 

Lets remind ourselves of the derivation of (\ref{deg N=8 sec Hecke}) in \cite{Maldacena:1999bp}. The index $d(Q,P)$ in (\ref{deg N=8 sec Hecke}) is the Fourier coefficient of the modified elliptic genus on the symmetric product orbifold CFT,
that is, the modified elliptic genus on $P^2/2$ copies of the $T^4$ CFT. The modified elliptic genus $\mathcal{E}''$ is defined as follows
\begin{equation}
\mathcal{E}''(q,y)\equiv \partial_{\bar{y}}^2Z\left(\text{Sym}^{P^2/2}(T^4);q,\bar{q},y,\bar{y}\right)|_{\bar{y}=1}.
\end{equation}
Here $y, \bar{y}$ couple respectively to the left and right fermion numbers, which are the R-symmetry charges. We take two derivatives to remove the two complex fermion zero modes the CFT carries; one can show that the modified elliptic genus is still invariant under deformations of the theory. 

The partition function for the symmetric product $Z(\text{Sym}^{P^2/2}(T^4))$ was derived using the DMVV formula \cite{Dijkgraaf:1996xw}. This formula can be written in terms of Hecke operators acting on the seed theory partition function, that is,
\begin{equation}
\sum_N p^N Z\left(\text{Sym}^N(\mathcal{M});q,\bar{q},y,\bar{y}\right)=\exp{\left[\sum_{m}p^m V_m Z_{1}(\mathcal{M},q,\bar{q},y,\bar{y})\right]},
\end{equation}
where the operator $V_m$ is the Hecke operator of index $m$ and weight zero acting on $Z_1$. The action of the Hecke operator $V_m$ on the Fourier coefficients of $Z_1(q,\bar{q},y,\bar{y})=\sum d(n,\bar{n},l,\bar{l})q^n\bar{q}^{\bar{n}}y^l\bar{y}^{\bar{l}}$ is
\begin{equation}\label{Hecke on Z1}
V_m \{d(n,\bar{n},l,\bar{l})\}=\sum_{s|(n-\bar{n},l,\bar{l},m)}s^{-1}d(nm/s^2,\bar{n}m/s^2,l/s,\bar{l}/s).
\end{equation}
Note that if $Z_1$ has zero weight under modular transformations, then $V_mZ_1$ is also modular invariant. However, if $Z_1$ is not modular invariant, but only transforms covariantly,  then $V_m Z_1$ does not have modular properties. To see this, let us define the action of a Hecke operator acting on a general modular object $f(\tau,\bar{\tau},z,\bar{z})$. Following \cite{Eichler:1985ja}, we have
\begin{equation}\label{Hecke non-holomorphic}
V_m f(\tau,\bar{\tau},z,\bar{z})=m^{\omega+\bar{\omega}-1}\sum_{\substack{\mu\in \Gamma_1\backslash M_m}} f|_{\substack{\omega,\bar{\omega}\\ k,\bar{k}}}(\mu),
\end{equation}
where $M_m$ is the set of $2\times 2$ matrices of determinant $m$ and $\Gamma_1\in PSL(2,\mathbb{Z})$. We have defined
\begin{equation}
f|_{\substack{\omega,\bar{\omega}\\ k,\bar{k}}}(\mu)=(c\tau+d)^{-\omega}(c\bar{\tau}+d)^{-\bar{\omega}}e^{-2\pi i m k \frac{cz^2}{c\tau+d}}e^{-2\pi i m \bar{k} \frac{c \bar{z}^2}{c\bar{\tau}+d}}f\left(\frac{a\tau+b}{c\tau+d},\frac{a\bar{\tau}+b}{c\bar{\tau}+d},\frac{z}{c\tau+d},\frac{\bar{z}}{c\bar{\tau}+d}\right),
\end{equation}
 with $k,\bar{k}$ the holomorphic and anti-holomorphic indices, and $\omega$ and $\bar{\omega}$ the weights of the holomorphic and anti-holomorphic sectors respectively. It is easy to see that (\ref{Hecke non-holomorphic}) is modular by construction. 
 
 We compute the Fourier coefficients of (\ref{Hecke non-holomorphic}). An element of $ M_m$ can always be written in the  form
\begin{equation}
\left(\begin{array}{cc}
a & b\\
c & d
\end{array}\right)=\left(\begin{array}{cc}
a' & b'\\
c' & d'
\end{array}\right)\left(\begin{array}{cc}
s & b''\\
0 & s'
\end{array}\right),\qquad \left(\begin{array}{cc}
a' & b'\\
c' & d'
\end{array}\right)\in PSL(2,\mathbb{Z}),
\end{equation}
with $ss'=m$ and $b'' \sim b'' \text{ mod}(s')$. The condition on $b''$ comes from the fact that the matrix $\left(\begin{array}{cc}
s & b''\\
0 & s'
\end{array}\right)$ is defined up to left multiplication by $\pm \left(\begin{array}{cc}
1 & r\\
0 & 1
\end{array}\right)\in PSL(2,\mathbb{Z})$ and $r\in \mathbb{Z}$. Inserting this back in (\ref{Hecke non-holomorphic}) one finds
\begin{equation}\label{sum s f}
V_m f(\tau,\bar{\tau},z,\bar{z})=\sum_{s,b''}s^{\omega+\bar{\omega}-1}f\left(\frac{s\tau+b''}{s'},\frac{s\bar{\tau}+b''}{s'},sz,s\bar{z}\right).
\end{equation}
If we write $f(\tau,\bar{\tau},z,\bar{z})=\sum_{n,\bar{n},l,\bar{l}}d(n,\bar{n},l,\bar{l})q^n\bar{q}^{\bar{n}}y^l\bar{y}^{\bar{l}}$, then the sum (\ref{sum s f}) translates into the following expression for the Fourier coefficients of $V_m f$ as
\begin{equation}\label{Tm Fourier coeff}
V_m\{c(n,\bar{n},l,\bar{l})\}=\sum_{s|(n-\bar{n},l,\bar{l},m)} s^{\omega+\bar{\omega}-1} d\left(mn/s^2,m\bar{n}/s^2,l/s,\bar{l}/s\right).
\end{equation}
The constraint on the divisor $s|(n-\bar{n},l,\bar{l},m)$ comes from the sum over $b''$.

Assuming that we have $\omega=\bar{\omega}$, we can conclude from the expression  (\ref{Tm Fourier coeff})  that  (\ref{Hecke on Z1}) are the Fourier coefficients of a modular object only when $Z_1$ is modular invariant, that is, when it has zero weight.  If $Z_1$ has  non-zero weight then we can reverse the steps (\ref{Hecke non-holomorphic}) and (\ref{sum s f}) presented above, starting with the coefficients (\ref{Hecke on Z1}). We find a formula similar to (\ref{Hecke non-holomorphic}), except that it comes with an anomalous term which forbids the expression of transforming correctly under modular transformations. As a corollary of this result, we find that the expression (\ref{deg N=8 sec Hecke}) can not be the Fourier coefficient of a modular object.  However, we can show that (\ref{deg N=8 sec Hecke}) can be written as the modified elliptic genus of an Hecke operator acting on the primitive answer. To be able to do this, one has to introduce  in the partition function the contribution of the four $U(1)$ currents, which arise due to the symmetries of the $T^4$. The inclusion of these currents renders the final answer modular invariant. The price to pay is that the partition function does not factorize anymore as a holomorphic times an anti-holomorphic function. In particular, we have
\begin{equation}
\tilde{Z}_1=\Theta_{U(1)^4}(\tau,\bar{\tau})\phi_{-2,1}(\tau,z)\bar{\phi}_{-2,1}(\bar{\tau},\bar{z}),
\end{equation}
where $\Theta_{U(1)^4}(\tau,\bar{\tau})$, which is a theta function, comes from the contribution of the four $U(1)$ currents.
 The action of $V_m$ on this function is now well defined and gives a modular invariant function. The Fourier coefficients $\tilde{c}(n,\bar{n},l)$ of the modified elliptic genus of $V_m \tilde{Z}_1$ are
\begin{eqnarray}
\tilde{c}(n,\bar{n},l)&=&\sum_{\bar{l}}\sum_{s|(n-\bar{n},l,\bar{l},m)}\bar{l}^2s^{-1}d(nm/s^2,\bar{n}m/s^2,l/s,\bar{l}/s)\nonumber\\
&=&\sum_{s|(n-\bar{n},l,m)}s\sum_{\bar{l}}\bar{l}^2d(nm/s^2,\bar{n}m/s^2,l/s,\bar{l}),
\end{eqnarray}
where in the second line we have interchanged the sums over $\bar{l}$ and $s$, and $d(n,\bar{n},l,\bar{l})$ are the Fourier coefficients of $\tilde{Z}_1$. Since we are not interested in the states that carry $U(1)^4$ quantum numbers, the sum over $\bar{l}$ projects over the sector with $\bar{n}=0$. The final result is therefore
\begin{equation}
\tilde{c}(n,\bar{n}=0,l)=\sum_{s|(n,l,m)} s\, c(nm/s^2,l/s),
\end{equation}
where $c(n,l)$ are the Fourier coefficients of $\phi_{-2,1}(\tau,z)$.

The main lesson to take from this exercise is that we can write the microscopic formula (\ref{deg N=8 sec Hecke}) in a form that is directly related to the action of a Hecke operator. Perhaps by understanding the physics related to the $U(1)^4$ currents, we can address the problem of why from the bulk theory we find a different power of $s$ in the non-primitive degeneracy formula.

\section{Discussion and Conclusion}

In this work, we have addressed arithmetic properties of black hole entropy in the context of the quantum entropy function. The main result is the application of arithmetic properties of generalized Kloosterman sums to the full $AdS_2$ partition function. To this end, the key results of our work are:

\begin{itemize}
	\item {\it{Sums of Kloosterman sums}}: we have develop and explored arithmetic properties of Kloosterman sums in the form of Selberg identities. We re-derived the Selberg identity of classical Kloosterman sums and extended those relations to the generalized version. A new object played a key role in this construction: a modified Kloosterman sum. These sums are based instead on two dimensional matrices with determinant greater than one. To our knowledge, these are novel properties of Kloosterman sums.
	
	\item  {\it{Arithmetics of quantum gravity}}: the path integral on the five dimensional $\mathbb{Z}_c$ orbifolds reproduces the Bessel function multiplied by a modified generalized Kloosterman sum. The Bessel function follows essentially from the analysis of \cite{Gomes:2017eac} and captures quantum fluctuations around the attractor background. The modified Kloosterman sums arise from the sum over flat connections in three dimensional Chern-Simons theory defined on the $M_{(c,d)}$ geometry. The key difference between this approach and the computations of \cite{Dabholkar:2014ema,Gomes:2017bpi} resides on the fact that we relax the condition that $(c,d)$ are co-prime. The map between the homology of the boundary cycles of $T^2$ and the full geometry $AdS_2\times S^1$ is characterized by fillings of the form $ad-bc=r$ with $r>1$. The Chern-Simons computation in this geometry leads directly to the modified version of the Kloosterman sums. Summing over all the orbifold geometries reproduces the structure of the non-primitive answers, including the dependence on the U-duality invariants in both the $\CN=8$ and $\CN=4$ examples. 
	
	\item {\it{$r\D6-r\aD6$ configurations and the proposal of \cite{Gomes:2017eac}}}: underlying the construction of the geometries $M_{(c,d)}$, with $(c,d)$ not necessarily co-prime, lies the proposal of \cite{Gomes:2017eac} for a first principles derivation of non-perturbative effects in the quantum entropy function, related to the polar states in the Rademacher expansion. The inclusion of the $r\D6-r\aD6$ configurations, or better their M-theory uplift with non-trivial fluxes, is key to the construction developed in this work. In particular, the inclusion of $r>1$ configurations is an essential ingredient in explaining why the full quantum entropy of a general  dyon reduces to a sum over the primitive degeneracies. Furthermore, the rank $r$ of the $\D6$ theory is related to the condition $ad-bc=r$ in the bulk theory.
	
\end{itemize}

Despite this success, we have not been able to account for the exact arithmetic function $g(s)$ that one finds in the microscopic formulas (\ref{non-primitive deg}). In both the $\CN=8$ and $\CN=4$ examples, such function is given by $g(s)=s$. However, our bulk computation predicts instead $g(s)=s^{\omega-1}$, where $\omega$ is the weight of the modular form. Our result ensures, nevertheless, that the full bulk degeneracy corresponds to a Fourier coefficient of a modular object, as expected from the holographic correspondence. This follows from the fact that for $g(s)=s^{\omega-1}$ we can see the degeneracy as the Fourier coefficient of a Hecke operator acting on the primitive answer, which ensures that the modular properties are preserved. In section \S \ref{sec Hecke}, we pointed out for  a possible solution to this puzzle. The discrepancy observed between the bulk and the microscopic degeneracies may be related to additional $U(1)$ currents in the CFT that have not been taken properly into account. We have shown that in order to obtain $g(s)=s$ from the symmetric product orbifold sigma-model, the additional currents play a very important role. From the quantum gravity point of view, one needs to understand how to include the charges dual to these currents in the computation of the Kloosterman sums. Perhaps this would render the necessary powers of $s$ to obtain $g(s)=s$. We leave this for future work.  

In this work, we claimed that in order to reproduce the arithmetic structure of the microscopic answers, we need to introduce in the path integral $AdS_2$ orbifolds with fixed points at the positions of the $\D6$ and $\aD6$. At this moment we do not fully understand  how the $\D6$ branes physically regulate such singularities.  It would be important to clarify this point. Another question concerns the computation of the Donaldson-Thomas invariants that count bound states of $\D6-\D4-\D2-\D0$ branes, for rank $r>1$ bundles. It is possible that such invariants are related to the arithmetic properties we considered here. It would be interesting to check this.

From the holography point of view, our results are a non-trivial test of the AdS/CFT correspondence at finite $"N"$, including non-perturbative phenomena. In the context of black holes, our results seem to connect quantum gravity and number theory at a deeper level.  Indeed, we have found that the structure of the non-perturbative answer, deeply rooted in the Kloosterman sums, is  tightly connected with the discreteness of quantum gravity. Along the way, this has allowed us to draw remarkable connections between Chern-Simons theory and number theory worth to be explored even further.

\subsection*{Acknowledgments}

This work is supported by Nederlandse Organisatie voor Wetenschappelijk Onderzoek (NWO) via a Vidi grant. We would like to thank the Delta ITP and the Institute for Theoretical Physics at the Utrecht University were this project was initiated, and also the Aspen Center for Physics, which is supported by the National Science Foundation grant PHY-1607611, where part of this work was completed.

\bibliographystyle{JHEP}
\bibliography{measure2}

\end{document}